\documentclass[aip,reprint,superscriptaddress]{revtex4-1}
\usepackage{amssymb,amsmath}
\usepackage[table]{xcolor}
\usepackage{graphicx}
\usepackage{algorithm}
\usepackage{algorithmicx}
\usepackage{algpseudocode}
\usepackage{url}

\newcommand{\E}[1]{\mathbb{E}\left[#1\right]}
\newcommand{\var}[1]{\text{var} \left[ #1 \right]}
\let\polishl\l
\renewcommand{\l}{\left}
\renewcommand{\r}{\right}
\newcommand{\subeq}[2]{\begin{subequations}\label{#2}\begin{align}#1\end{align}\end{subequations}}
\newcommand{\eq}[2]{\begin{equation}#1\label{#2}\end{equation}}
\newcommand{\al}[2]{\begin{align}#1\label{#2}\end{align}}

\newcommand{\red}[1]{{\color{black}{#1}}}

\newcommand{\dd}{\text{d}}

\usepackage{CJKutf8}
\begin{document}

\title{Scaling methods for accelerating kinetic Monte Carlo simulations of chemical reaction networks}
\author{Yen Ting Lin}
\email{Electronic mail: yentingl@lanl.gov.}
\affiliation{Center for Nonlinear Studies and Theoretical Biology and Biophysics Group, Theoretical Division, Los Alamos National Laboratory, New Mexico 87545, USA}
\affiliation{Yen Ting Lin and Song Feng contributed equally to this work.}
\author{Song Feng (\begin{CJK*}{UTF8}{gbsn}冯松\end{CJK*})}
\email{Electronic mail: song.feng@outlook.com.}
\affiliation{Center for Nonlinear Studies and Theoretical Biology and Biophysics Group, Theoretical Division, Los Alamos National Laboratory, New Mexico 87545, USA}
\affiliation{Yen Ting Lin and Song Feng contributed equally to this work.}
\author{William S. Hlavacek}
\email{Electronic mail: wish@lanl.gov.}
\affiliation{Center for Nonlinear Studies and Theoretical Biology and Biophysics Group, Theoretical Division, Los Alamos National Laboratory, New Mexico 87545, USA}

\date{\today}
\begin{abstract}
Various kinetic Monte Carlo algorithms become inefficient when some of the population sizes in a system are large, which gives rise to a large number of reaction events per unit time. Here, we present a new acceleration algorithm based on adaptive and heterogeneous scaling of reaction rates and stoichiometric coefficients. The algorithm is conceptually related to the commonly used idea of accelerating a stochastic simulation by considering a sub-volume $\lambda \Omega$ ($0<\lambda<1$) within a system of interest, which reduces the number of reaction events per unit time occurring in a simulation by a factor $1/\lambda$ at the cost of greater error in unbiased estimates of first moments and biased overestimates of second moments. Our new approach offers two unique benefits. First, scaling is adaptive and heterogeneous, which eliminates the pitfall of overaggressive scaling. Second, there is no need for an \emph{a priori} classification of populations as discrete or continuous (as in a hybrid method), which is problematic when discreteness of a chemical species changes during a simulation. The method requires specification of only a single algorithmic parameter, $N_c$, a global critical population size above which populations are effectively scaled down to increase simulation efficiency. The method, which we term partial scaling, is implemented in the open-source BioNetGen software package. We demonstrate that partial scaling can significantly accelerate simulations without significant loss of accuracy for several published models of biological systems. These models characterize activation of the mitogen-activated protein kinase ERK, prion protein aggregation, and T-cell receptor signaling.
\end{abstract}

\maketitle

\section{Introduction}

Kinetic Monte Carlo (KMC) procedures \cite{voter2007introduction}, such as the well-known direct and next-reaction methods of Gillespie\cite{Gillespie2007}, are commonly used to study stochastic chemical kinetics, especially in biochemical systems. These procedures have been widely used to study systems in which population fluctuations arise from intrinsic molecular noise and the discreteness of the chemical species of interest \cite{McAdams1999}. Population fluctuations are characteristic of many genetic regulatory circuits, as populations in these systems tend to be discrete.

KMC procedures are also useful for studying systems, such as cell signaling networks, that have large state spaces \cite{Suderman2018} arising from the combinatorial number of chemical species that can be generated by biomolecular interactions of interest \cite{Chylek2014}. For such systems, it may be impracticable, even with the aid of a computer, to enumerate the chemical species that are potentially populated. Nonetheless, simulations can be performed by formulating rules to represent biomolecular interactions\cite{Chylek2014} and then using these rules as event generators in a so-called network-free simulation algorithm \cite{Suderman2018}, such as that implemented in the NFsim software package \cite{Yang2008,Sneddon2011}. In a network-free simulation (of cell signaling dynamics), stochastic effects may be negligible \cite{Creamer2012}. Populations in cell signaling networks tend to be large, such that population densities are nearly continuous variables.

In any exact KMC procedure, system state is updated (and time is advanced) only one reaction event at a time, which can be a severe limitation when the number of events per unit time is large. In many applications of KMC aimed at studying noisy system behavior, this difficulty is not encountered because populations of interest are small, which makes for efficient simulation. However, inefficiency may arise if some population sizes are large and/or some reactions are fast. Both of these features of a reaction system can introduce a large, unmanageable number of reaction events per unit time. A variety of approaches have therefore been developed for addressing this problem.

In the near-continuum limit where populations of chemical species are uniformly large (but not infinite, such that concentrations can be appropriately treated as continuous variables), the diffusive stochastic differential equation (SDE) obtained from the diffusion approximation \cite{van1992stochastic,risken1996fokker} provides an efficient solution for simulating stochastic chemical kinetics. Gillespie's $\tau$-leaping method\cite{gillespie2001approximate} provides an efficient solution for another regime of behavior, where the discreteness of populations is still relevant. In this method, a time window $\tau$ is prescribed. Importantly, $\tau$ should be (1) large enough such that multiple events occur within this window and (2) small enough such that the system's configuration and state transition rates (also called reaction propensity functions) do not change significantly as a result of the events occurring within the window. Unfortunately, these opposing requirements limit applicability of the method. Finally, hybrid methods have been developed and applied to solve specific problems\cite{Salis05accurate,newby2013breakdown,bokes2013transcriptional,bressloff2014stochastic,bressloff2015path,lin2016gene,lin2016bursting,bressloff2017stochastic,bressloff2017feynman,lin2018efficient}. A characteristic feature of these methods is the division of state variables into two distinct sets, the discrete variables and the continuous variables. The continuous variables are evolved forward in time via numerical integration of ordinary differential equations (ODEs) while the discrete variables are simultaneously evolved forward in time via KMC. A general framework for hybrid simulation is provided by the formalism of discretely switching Langevin dynamics\cite{mao2006stochastic}.

All of the above-mentioned methods are designed to trade accuracy for efficiency. In other words, acceleration gains are obtained through simplifying assumptions that introduce approximations. The diffusive SDE approach is based on the assumption that only the first and second moments of distributions are important; higher order moments are ignored. The $\tau$-leaping approach is based on the assumption of Poissonian statistics in the time window $\tau$. Hybrid approaches are based on the assumption that fluctuations in certain populations can be ignored entirely over the full time window of simulation. For most, if not all, of these methods, there is no way to determine \emph{a priori} if their approximations are tolerable, and unacceptable errors can arise unexpectedly during the course of a simulation. SDE and hybrid approaches become inappropriate if populations of all or a subset of chemical species, respectively, are not sufficiently large throughout a simulation. As we will see, for some systems, population sizes are distributed across multiple scales, and the size of a population can change qualitatively (e.g., from large to discrete to large again) during the course of a simulation, as when a system exhibits oscillatory behavior. The $\tau$-leaping method is also problematic: for a given system, there is no guarantee that a time window $\tau$ having the necessary properties exists. For all of these methods, approximation accuracy depends in a non-obvious way on parameters (initial conditions and rate constants). Thus, their use in any analysis involving variation of parameter values, such as a curve fitting procedure, requires careful consideration.

An alternative, simpler idea for accelerating stochastic simulation is to consider only a sub-volume of the system of interest, taking this sub-volume to be representative of the whole system. Given a system with volume $\Omega$, if we simulate reactions in only a sub-volume $\lambda \Omega$, where $0 < \lambda < 1$, we can expect to reduce simulation costs by the factor $1/\lambda$ because the number of reaction events per unit time is reduced by this factor. As we will discuss later in detail, this homogeneous and static scaling approach can speed up estimation of first moments of the stochastic dynamics, and estimates are unbiased so long as populations are not scaled too aggressively to size of order 1 or below. A drawback of scaling is that, with achievement of any acceleration, there is an unavoidable bias in estimation of higher moments. Another drawback is that the acceleration attainable with reasonable accuracy may be limited by the small population size of one or more critical chemical species. If system behavior is influenced by a chemical species with a small population size, as in stochastic switching\cite{Kepler2001}, the setting for $\lambda$ cannot be such that the population of the chemical species falls near or below 1, which would fundamentally change or eliminate its influence on behavior.

Scaling is commonly used as part of a (network-free) stochastic simulation approach when the dynamics of interest are nearly deterministic, as recommended, for example, by Faeder et al.\cite{Faeder2009}. In such cases, the goal of simulation is simply to calculate first moments. The inability of scaling to yield unbiased estimates of higher moments is of little concern. However, in these applications, there remains a need to avoid overaggressive scaling of critical populations. As we will see through examples, this constraint can severely limit the usefulness of scaling via the standard (homogeneous and static) approach.

Here, we present a new approach for accelerating stochastic simulation through scaling, which we term partial scaling. In partial scaling, for each enumerated individual reaction considered in a system, its reaction rate is scaled by a factor that effectively makes the smallest reactant or product population, if greater than a critical population size $N_c$, equivalent or close to this critical population size. Stoichiometric coefficients are scaled by the inverse of the scaling factor for reaction rate. No scaling is applied if the smallest reactant or product population is already at or below $N_c$. The scaling is heterogeneous (vs. homogeneous) because each reaction has its own scaling factor, which is determined by $N_c$ and the smallest population of the chemical species involved in the reaction. Furthermore, reaction rates are scaled \emph{on the fly} as population sizes change over time, i.e., scaling is adaptive (vs. static). Because scaling is not performed for a reaction when the smallest population of the participating chemical species is smaller than $N_c$, partial scaling permits not only unbiased estimation of first moments but also less biased estimation of second moments compared to the standard scaling method. Although partial scaling entails heterogeneous and adaptive scaling, whereas standard scaling entails homogeneous and static scaling, both methods involve only a single algorithmic parameter: $\lambda$ in the case of standard scaling and $N_c$ in the case of partial scaling.

The remainder of this report is organized as follows. In Sec.~II, we introduce notation. In Sec.~III, we review how scaling is currently used to accelerate stochastic simulation. Through analysis of a simple reaction network, we explain why this approach yields unbiased estimates of first moments while improving simulation efficiency at the expense of biased estimates of higher moments (e.g., overestimates of variance). We also illustrate the consequences of overaggressive scaling. In Sec.~IV, to address the limitations of standard scaling, we introduce partial scaling. To complement the description of the method, we provide pseudocode for a partial scaling algorithm. In Sec.~V, using published models for three complex biochemical reaction networks\cite{kochanczyk2017relaxation,rubenstein2007dynamics,lipniacki2008stochastic}, we evaluate the performance of partial scaling relative to standard scaling. In Sec.~VI, we call attention to our general-purpose implementation of partial scaling in the BioNetGen software package\cite{Harris2016}. We conclude with a discussion of results and possible future directions in Sec.~VII.

\section{Notation}
We consider a closed, well-mixed, and isothermal reaction system having constant volume\footnote{The parameter $\Omega$ characterizes system size. Here, we take it to represent volume but it can alternatively be interpreted as a population that defines a population scale. With this interpretation, (dimensionless) population densities replace concentrations as the state variables in the continuum limit with no change in the mathematical form of the governing equations.} $\Omega$. The system contains a (dilute) mixture of up to $M$ chemical species, $X_i,\ldots,X_M$, that react within a network of $R$ reactions. We assume that each reaction $r\in\{1,\ldots,R\}$ obeys mass-action kinetics with rate constant $\kappa_r$. We use $N_i$ to denote the non-negative and discrete population of $X_i$. The discrete state of the system is defined by the vector $\mathbf{N} \equiv \l(N_1, \ldots, N_M \r)$. We use $\mathbb{P}_{\mathbf{N}}\l(t\r)$ to denote the probability that the system is in state $\mathbf{N}$ at time $t$. In the continuum limit, reached as $\Omega \rightarrow \infty$ and $N_i \rightarrow \infty$ while $N_i/\Omega$ remains constant for all $i$, the concentration of each $X_i$, defined as $n_i \equiv N_i/\Omega$, is a continuous variable, and system state is defined by the vector $\mathbf{n} \equiv (n_1, \ldots, n_M)$. We use $\rho\l(\mathbf{n}, t\r)$ to denote the probability density of state $\mathbf{n}$ at time $t$.

\section{Static and homogeneous scaling}
In this section, which can be bypassed by readers who are expert in stochastic modeling, we provide a brief review of scaling as a tool to accelerate stochastic simulation, mainly through the exercise of analyzing a simple reaction system. Our goal is to introduce background facts important for appreciating the benefits and limitations of scaling, which should strictly be considered only for a system with large populations, such that (all) concentrations are nearly continuous. As we will see later, many biochemical systems have concentrations that are nearly continuous. The background facts of concern here are as follows. First, the number of events generated per unit time in an exact stochastic simulation is an extensive quantity. Accordingly, the efficiency of an exact stochastic simulation algorithm (SSA) scales with system size. Second, if after scaling, the system of interest remains near the continuum limit, exact simulations of the scaled system yield unbiased estimates of the first moments of the random variables that are being sampled in the simulations. Third, scaling can introduce discreteness in populations that is inappropriate. In other words, overaggressive scaling can introduce systematic errors.

An exact SSA generates sample paths of the stochastic dynamics of a (well-mixed) reaction system by executing a series of reaction events (in a Monte Carlo fashion) that each brings about discrete changes of the populations of the reacting chemical species. State transition rates are characterized by propensity functions that, by construction, are extensive quantities \cite{gillespie1977exact,van1992stochastic}. Thus, as we will see shortly for a specific example, the values of these functions scale linearly with system size $\Omega$, and as $\Omega$ increases, an SSA generates more events per unit time. Events occur at an overall rate proportional to $\Omega$ and inversely proportional to the average waiting time between events. Consequently, as $\Omega$ increases, execution of an SSA eventually becomes too computationally expensive to be practical.

When system size is large, the joint probability distribution of the discrete-state random process sampled by an SSA can be approximated by various methods. Ordered by degree of granularity of approximation, these methods include (1) $\tau$-leaping, which approximates the state-dependent transition kernel of an SSA by a constant (Poissonian) transition kernel, which is used over a small time window $\tau$\cite{gillespie2001approximate,cao2005avoiding,cao2006efficient}; (2) the diffusion approximation, which coarse-grains the discrete state space into a continuous state space via the Kramers--Moyal expansion, the end result of which is a Fokker--Planck equation describing the joint probability density of the system’s state characterized in terms of intensive state variables, such as concentrations or population densities\cite{kramers1940brownian,moyal1949stochastic,van1992stochastic,gardiner1985handbook,doering2005extinction,lin2012features,lin2012features,lin2015demographic,lin2015demographic2}; and (3) the linear-noise approximation (LNA), which linearizes the multiplicative noise kernel of a Fokker--Planck equation, such that the resulting dynamics involves only Gaussian white noise\cite{kampen1961power,van1992stochastic}. 

Another approach for coping with large system size is to use an SSA to simulate the stochastic dynamics not in the whole system of interest but rather in a suitably small sub-volume $\lambda\Omega$, where $0<\lambda<1$. The dynamics of the sub-volume are taken to represent the dynamics of the whole system. This approach, which we will refer to as \textit{static and homogeneous scaling} or standard scaling, is seemingly justified for a well-mixed system, and from our discussion above, it can be expected to reduce the cost of simulation by the factor $1/\lambda$: the smaller the value of $\lambda$, the greater the efficiency gain. However, the approach can produce erroneous and misleading results if scaling is overaggressive, as when scaling reduces the population of a critical chemical species (i.e., one that influences overall system behavior even if its population size is relatively small) to 0 or $\mathcal{O}(1)$. In these cases, the dynamics in the sub-volume will differ qualitatively from the dynamics in the whole system. For example, scaling to a population size $\sim 1$ will introduce bursty behavior, which has unique statistical properties\cite{lin2016gene,lin2016bursting,lin2018efficient}. Thus, scaling should be limited, but it is not clear how to impose an appropriate lower bound on $\lambda$. A constraint such as $\lambda \ge \min_{1 \le i \le M} N_i \l(t\r)$ requires a trial-and-error procedure to select $\lambda$ because the bounding term is dependent on both time and parameters (initial conditions and rate constants). Another concern, even with an appropriate setting for $\lambda$, is how scaling affects the statistics of sample paths. As we will see, estimates for first moments are unbiased but scaling leads to biased overestimates of second moments.

\begin{figure*}[t]
\begin{center}
\includegraphics[width=1.0\textwidth]{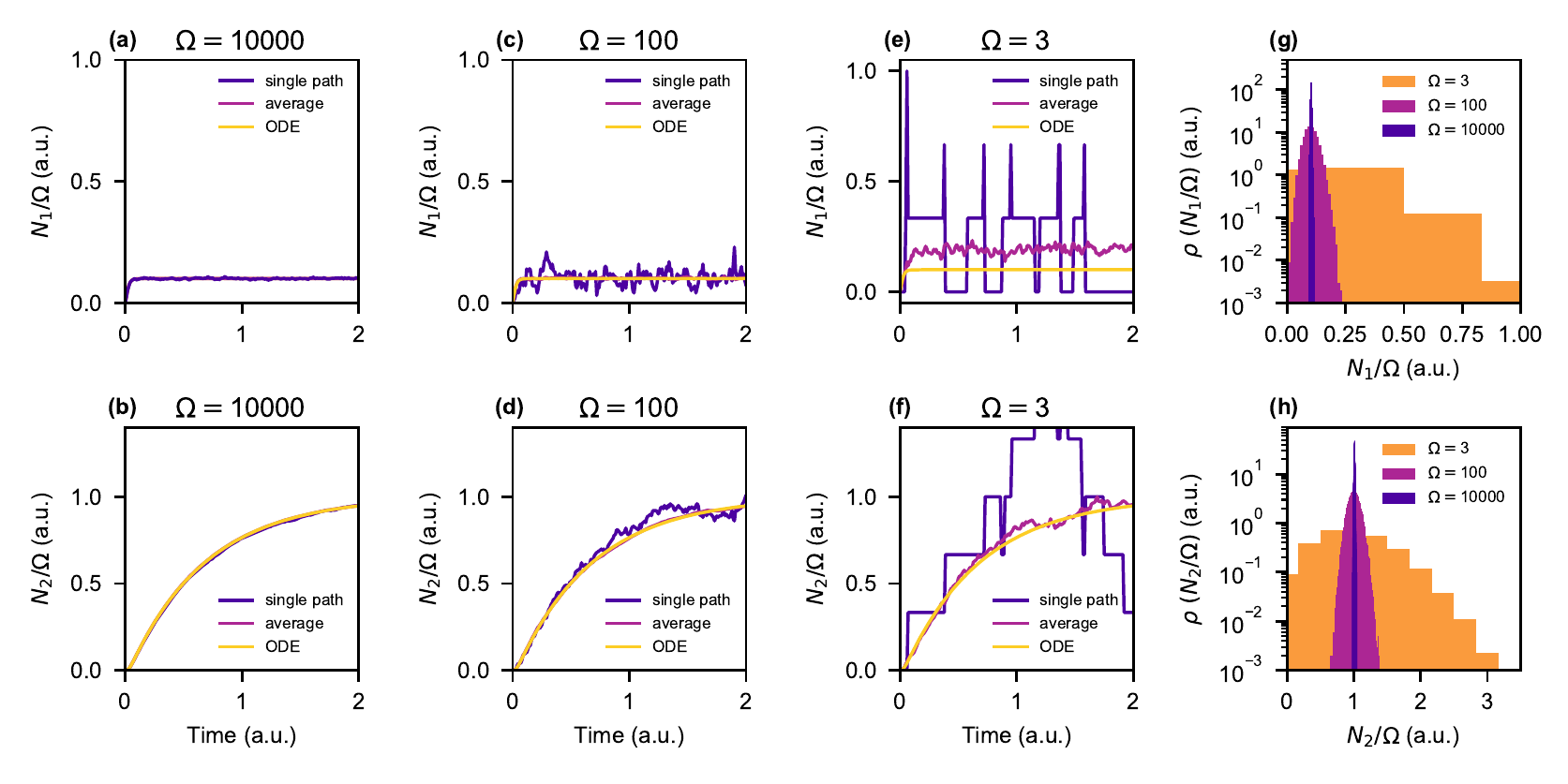}
\caption{Simulations based on Eqs.~\eqref{eq:toyModel}. In the six panels at left (a--f), we compare stochastic sample paths found using an SSA to the corresponding deterministic trajectories found by numerically integrating the ODEs in Eq.~\eqref{eq:rateEq}. We consider a range of system sizes (without changing intensive parameters): $\Omega=10^4$ (a and b), $10^2$ (c and d), and $3$ (e and f). The model parameters are $\kappa_1=3$, $\kappa_2=300$, and $\kappa_3=1.5$. In the top panels (a, c and e), we compare the random variable $N_1/\Omega$ to the deterministic quantity $n_1$. Similarly, in the bottom panels (b, d and f), we compare $N_2/\Omega$ to $n_2$. For panels (a--f), as indicated by the legend of each panel, we plot a single stochastic sample path, the first moment estimated by averaging over 500 sample paths, and the corresponding deterministic trajectory. In panels (g and h), we plot the marginal stationary distributions for $N_1/\Omega$ and $N_2/\Omega$, each measured from $10^5$ sample paths. Note that, when $\Omega=3$, the probability distribution of $N_1/\Omega$ is skewed to the right (g), leading to a first moment estimation that deviates noticeably from the continuum limit (e). Time and volume each have arbitrary units (a.u.).}
\label{fig:dimerization}
\end{center}
\end{figure*}

\subsection{A simple reaction network}\label{sec:toy-model}
To illustrate features of static and homogeneous scaling, we will consider the following reaction network, consisting of zeroth-, first-, and second-order elementary reactions:
\subeq{
\varnothing \xrightarrow{\kappa_1}{}&  X_1, \label{eq:reaction1}  \\
X_1+X_1\xrightarrow{\kappa_2}{}& {X_2},  \label{eq:reaction2} \\
X_2 \xrightarrow {\kappa_3}{}& \varnothing .  \label{eq:reaction3} 
}{eq:toyModel}
Here, $M=2$ and $R=3$. The chemical species $X_1$ is injected via a zeroth-order reaction into the system (volume $\Omega$) with rate constant $\kappa_1$. In a second-order reaction, two copies of $X_1$ react irreversibly to form $X_2$ with rate constant $\kappa_2$. The chemical species $X_2$ is removed from the system via a first-order reaction with rate constant $\kappa_3$.

\subsection{Deterministic and stochastic chemical kinetics}\label{sec:modelConstruction}

For the reaction network of Eqs.~\eqref{eq:toyModel}, we can write the following ODEs for mass-action kinetics:
\subeq{
\frac{\dd n_1}{\dd t}  ={}& \kappa_1 -  \kappa_2  n_1^2, \label{eq:rateEq1}\\
\frac{\dd n_2}{\dd t}  ={}& \frac{1}{2} \kappa_2  n_1^2 - \kappa_3 n_2\label{eq:rateEq2},
}{eq:rateEq} 
where $n_1(t)$ is the time-dependent concentration of $X_1$ and $n_2(t)$ is the time-dependent concentration of $X_2$. It should be noted that the $1/2$ factor in Eq.~\ref{eq:rateEq2} accounts for the symmetry of the left-hand side of Eq.~\eqref{eq:reaction2}\footnote{The symmetry factor is not strictly required in the context of ODE modeling of chemical kinetics (because a constant times a constant is still a constant), but it emerges from the CME description of chemical kinetics.} and that all terms on the right-hand side of Eqs.~\eqref{eq:rateEq} incorporate the stoichiometric coefficients of the reactions in Eqs.~\eqref{eq:toyModel}. The terms $\kappa_1$, $\kappa_1n_1^2$, $(1/2)\kappa_2n_1^2$, and $\kappa_3n_2$ are the mass-action rate laws that follow from reactions $r=1,2,3$ in Eqs.~\eqref{eq:toyModel}; these terms give the rates of $X_1$ injection, $X_1$ consumption, $X_2$ generation, and $X_2$ removal, respectively. The ODEs of Eqs.~\eqref{eq:rateEq} provide a deterministic description of the chemical kinetics, with each $n_i$ corresponding to $N_i/\Omega$ in the continuum limit.

To obtain a stochastic description of the chemical kinetics \cite{gillespie1977exact,van1992stochastic}, let us enumerate the possible state transitions when the system is in state $\mathbf{N} = \l(N_1,N_2\r)$. From Eqs.~\eqref{eq:toyModel}, the possible state transitions and their stochastic transition rates (given above the arrows) are as follows:
\subeq{
(N_1, N_2) \xrightarrow{h_1(\mathbf{N})\times\bar{\kappa}_1}{}& (N_1+1, N_2),  \label{eq:sto1}\\
(N_1, N_2) \xrightarrow{h_2(\mathbf{N})\times\bar{\kappa}_2}{}& (N_1-2, N_2+1), \label{eq:sto2}\\ 
(N_1, N_2) \xrightarrow{h_3(\mathbf{N})\times\bar{\kappa}_3}{}& (N_1, N_2-1). \label{eq:sto3}
}{eq:sto}
Here, we use $h_r\l(\mathbf{N}\r)$ to denote the number of ways reaction $r$ can take place and $\bar{\kappa}_r$ to denote the transition rate for reaction $r$ when this reaction can take place in one and only one way\footnote{The limit of $h_r\bar{\kappa}_r\Delta t$, evaluated at time $t$, as $\Delta t \rightarrow 0$ equals the probability that a reaction $r$ takes place somewhere in the system within a time window of $t$ to $t+\Delta t$. A good approximation of the probability is obtained with finite $\Delta t$ so long as  $\Delta t$ is small enough such that the probability of two or more reactions of any kind occurring within the time window $[t,t+\Delta t)$ is much smaller than the probability of just one reaction.}. For reaction $r=1$ (Eq.~\eqref{eq:reaction1}), there is only one way to inject a copy of $X_1$ into the system. Thus, $h_1=1$. For reaction $r=2$ (Eq.~\eqref{eq:reaction2}), the number of ways that two copies of $X_1$ can react to form $X_2$ is $N_1$ choose 2. Thus, $h_2=C^{N_1}_2=N_1(N_1-1)/2$. Finally, for reaction $r=3$ (Eq.~\eqref{eq:reaction3}), there are $N_2$ choose 1 ways to remove a copy of $X_2$ from the system. Thus, $h_3=C^{N_2}_1=N_2$.

The transition rates in Eqs.~\eqref{eq:sto} are related to the mass-action rate laws of Eq.~\eqref{eq:rateEq}. The relationships are revealed by considering the continuum limit, where Eqs.~\eqref{eq:sto} must be consistent with Eqs.~\eqref{eq:rateEq}. Eq.~\eqref{eq:sto1} indicates that injection of $X_1$ increases the concentration of $X_1$ (i.e., $N_1/\Omega$), on average, by $\bar{\kappa}_1/\Omega$ per unit time, which is the rate characterized by the rate law $\kappa_1$ in Eq.~\eqref{eq:rateEq1}. Thus, we find that $\bar{\kappa}_1=\kappa_1\Omega$. Eq.~\eqref{eq:sto2} indicates that conversion of $X_1$ to $X_2$ decreases the concentration of $X_1$ (i.e., $N_1/\Omega$), on average, by $2\times (1/2)\bar{\kappa}_2N_1(N_1-1)/\Omega$ per unit time (noting that each firing of the reaction in Eq.~\eqref{eq:sto2} consumes two copies of $X_1$), which is the rate characterized by the rate law $\kappa_2 n_1^2$ in Eq.~\eqref{eq:rateEq1}. Thus, taking $N_1 \ll N_1^2$, which is appropriate in the continuum limit, we find that $\bar{\kappa}_2=\kappa_2/\Omega$. Eq.~\eqref{eq:sto3} indicates that removal of $X_2$ decreases the concentration of $X_2$ (i.e., $N_2/\Omega$), on average, by $\bar{\kappa}_3N_2/\Omega$ per unit time, which is the rate characterized by the rate law $\kappa_3n_2$ in Eq.~\eqref{eq:rateEq2}. Thus, we find that $\bar{\kappa}_3=\kappa_3$. In general\cite{van1992stochastic}, for a $k^\text{th}$ order reaction $r$, $\bar{\kappa}_r$ is equal to $\kappa_r \Omega^{1-k}$.

\begin{figure*}[t]
\begin{center}
\includegraphics[width=1.0\textwidth]{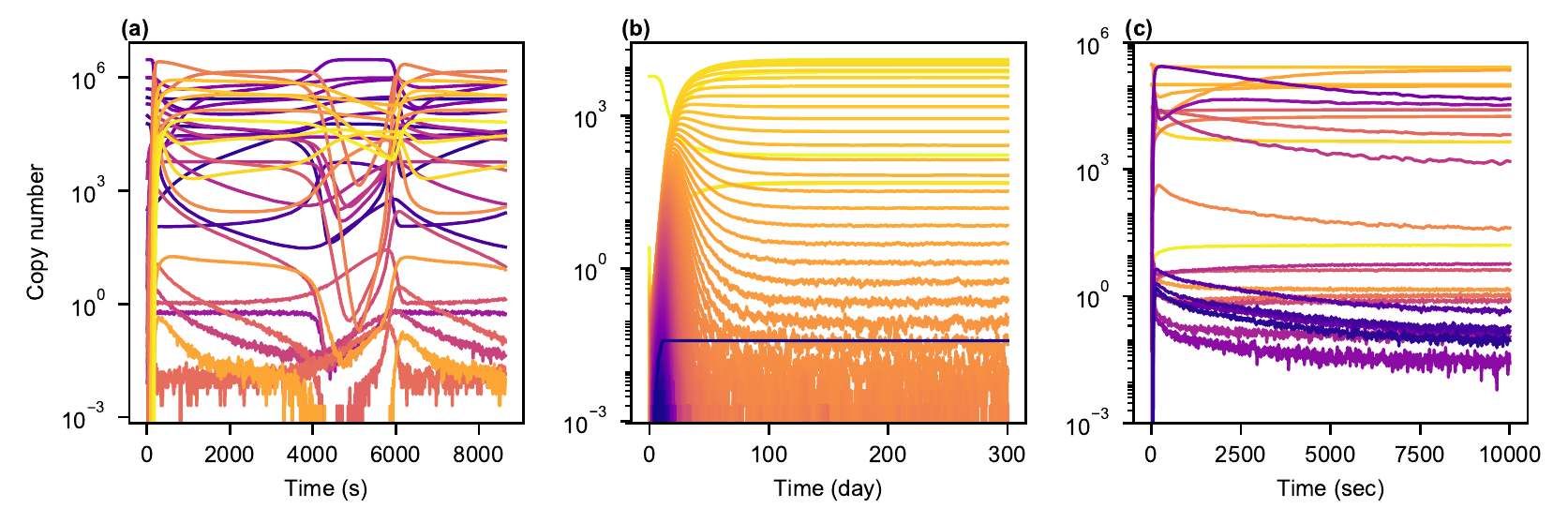}
\end{center}
\caption{The stochastic dynamics of models for (a) activation of the mitogen-activated protein kinase (MAPK) ERK\cite{kochanczyk2017relaxation}, (b) prion protein (PrP$^{\rm Sc}$) aggregation\cite{rubenstein2007dynamics}, and (c) T-cell receptor (TCR) signaling\cite{lipniacki2008stochastic}. 500 independent sample paths were generated through continuous-time Markov chain simulation\cite{schwartz2008biological,gillespie1977exact}, and the first moments of all chemical species considered in each model were each estimated by averaging over the generated sample paths. Plots in each panel show these first moment estimates. As can be seen, for each model, there is a broad spectrum of population scales, which spans several orders of magnitude. The models and simulations considered here are defined in BioNetGen input files (also called BNGL files) available online within the RuleHub repository \cite{ERK_BNGL,Prion_BNGL,TCR_BNGL}. The BNGL file for the ERK activation model\cite{ERK_BNGL} is an edited version of the BNGL file provided by Kocha{\'{n}}czyk et al.\cite{kochanczyk2017relaxation}. The BNGL file for the prion protein aggregation model\cite{Prion_BNGL} is new; we confirmed that this formulation of the model is consistent with the simulation results reported by Rubenstein et al.\cite{rubenstein2007dynamics}. The BNGL file for the TCR signaling model\cite{TCR_BNGL} is new. We rewrote the BNGL file originally provided by Lipniacki et al.\cite{lipniacki2008stochastic} for compatibility with current BNGL conventions; we confirmed that the new formulation of the model is consistent with the original formulation.}
\label{fig:models}
\end{figure*}

\subsection{Efficiency, accuracy and precision of standard scaling}\label{sec:uniformScalingAnalysis}

The utility of scaling derives from the fact that transition rates (i.e., propensity functions) are extensive, meaning that they scale with system size. To verify this claim, consider the continuum limit, which is approached as $\Omega \rightarrow \infty$ and $N_i \rightarrow \infty$ for all $i$, but with each $N_i/\Omega$ held constant. Thus, in the finite but large population limit, $N_i \propto \Omega$ $\forall i$. Accordingly, the stochastic transition rate of any $k^\text{th}$ order reaction $r$ has magnitude given by the expression $\bar{\kappa}_r \mathcal{O}(\Omega^k)$. However, from the relationship discussed above, $\bar{\kappa}_r \propto \kappa_r \Omega^{1-k}$, we can see that the expression simplifies to $\mathcal{O}(\Omega)$, indicating that the number of reaction events per unit time is proportional to the system size. By reducing system size $\Omega \rightarrow \lambda \Omega$, a speed up by a factor of $1/\lambda$ can be obtained as events occur $1/\lambda$ times less frequently.

A key idea of scaling is to take the stochastic dynamics within in a sub-volume of a system, or scaled system, to be representative of the dynamics in the whole system, or unscaled system. What this means in practice is as follows. In simulation of a scaled system, when a reaction event causes a change in the population of species $X_i$ by $\Delta N_i$, we take this change to correspond to a change of $\Delta N_i/\lambda$ in the unscaled state space. Thus, the method of static and homogeneous scaling can be seen as scaling down the values of all stochastic transition rates by a factor $\lambda$ and scaling up the values of all stoichiometric coefficients by a factor $1/\lambda$. As we will see later, this interpretation can be generalized to make the scaling of transition rates and stoichiometric coefficients adaptive and heterogeneous. However, let us first establish the accuracy and precision of static and homogeneous scaling, which we will pursue through an analysis of the reaction network of Eqs.~\eqref{eq:toyModel}.

We start by considering the chemical master equation\cite{gardiner1985handbook,van1992stochastic} (CME) for the reaction network of Eqs.~\eqref{eq:toyModel}. For arbitrary $N_1$ and $N_2$, the CME includes an ODE of the following form: 
\al{
\frac{\dd}{\dd t}  \mathbb{P}_{N_1,N_2} ={}&  - \Omega \kappa_1 \l[\mathbb{P}_{N_1,N_2} - \mathbb{P}_{N_1-1,N_2} \r]\nonumber \\
{}& -  \frac{\kappa_2}{\Omega}  \frac{N_1\l(N_1-1\r)}{2} \mathbb{P}_{N_1,N_2} \nonumber \\
{}& +  \frac{\kappa_2}{\Omega} \frac{\l(N_1+2\r)\l(N_1+1\r)}{2}\mathbb{P}_{N_1+2,N_2-1} \nonumber \\
{}& -  \kappa_3 \l[ N_2 \mathbb{P}_{N_1,N_2} -  \l(N_2+1\r)\mathbb{P}_{N_1,N_2+1}\r].  
}{eq:CME}
Recall that $\mathbb{P}_{\mathbf{N}}\l(t\r)$, written here as $\mathbb{P}_{N_1,N_2}$, denotes the probability of system state $\mathbf{N}=(N_1,N_2)$ at time $t$.

If the system has large populations (i.e., the probability mass concentrates in regions where both $N_1$ and $N_2$ are large), we can perform the Kramers--Moyal expansion\cite{kramers1940brownian,moyal1949stochastic,van1992stochastic} to transform the CME into a partial differential equation describing a diffusive process. This procedure begins by replacing extensive variables (discrete populations) by intensive ones (concentrations/population densities). We introduce the intensive variables $n_1 \equiv N_1/\Omega$ and $n_2 \equiv N_2/\Omega$. We also introduce $\dd n \equiv \dd n_1 \equiv \dd n_2 \equiv 1/\Omega$, which is the size of the grid in the state space of $n_1$ and $n_2$. Finally, we replace discrete probabilities with probability densities such that $\mathbb{P}_{N_1,N_2}(t)= \rho(n_1,n_2,t) \dd n_1 \dd n_2 = \rho(n_1,n_2,t) / \Omega^2$. Thus, after changing variables, we obtain
\al{
{}& \partial_t \rho \l(n_1,n_2,t\r) =- \Omega \kappa_1 \l[ \rho\l(n_1,n_2,t\r) - \rho\l(n_1-\dd n,n_2,t\r) \r]\nonumber \\
{}& - \kappa_2\Omega  \frac{n_1\l(n_1-\dd n\r)}{2} \rho\l(n_1,n_2,t\r) \nonumber \\
{}& + \kappa_2 \Omega \frac{\l(n_1+2\dd n\r)\l(n_1+\dd n\r)}{2} \rho \l(n_1+2\dd n,n_2-\dd n,t\r) \nonumber \\
{}& - \kappa_3 \Omega \l[ n_2 \rho\l(n_1,n_2,t\r) - \l(n_2 +\dd n \r) \rho\l(n_1,n_2+\dd n,t\r)\r].  
}{eq:CMEc}
At this stage, Eq.~\eqref{eq:CMEc} is simply a rewritten form of Eq.~\eqref{eq:CME}, obtained after changing from extensive to intensive variables. As such, $\rho(n_1,n_2,t)$ can only be evaluated on a lattice $(n_1,n_2)\in \l\{0, 1/\Omega, 2/\Omega, \ldots\r\}^2$.

The first key idea of the Kramers--Moyal expansion for the CME is to extend the domain of the function $\rho$ from $(n_1,n_2)\in \l\{0, 1/\Omega, 2/\Omega, \ldots\r\}^2$ to $\mathbb{R}_{\ge 0}^2$ while assuming \eqref{eq:CMEc} holds true for any \emph{non-negative and real-valued} $n_1$ and $n_2$. This transforms Eq.~\eqref{eq:CMEc} into a difference equation for $\rho$. The second key idea is that under the assumption that $\Omega\gg 1$, the \emph{difference equation} can be approximated by local derivatives.

Under the assumption that $\Omega \gg 1$ and $\dd n \ll 1$ in Eq.~\eqref{eq:CMEc}, which is viewed now as a difference equation, the Kramers--Moyal expansion procedure continues by Taylor expanding the terms $\rho\l(n_1 \pm \dd n, n_2 \pm \dd n\r)$ in Eq.~\eqref{eq:CMEc}. After all terms in each Taylor expansion higher than second order are dropped\footnote{Terms higher than second order are dropped because of the Pawula theorem\cite{risken1996fokker}, which states that any higher-order truncation ($\mathcal{O}\l(1/\Omega^2\r)$) fails to preserve the positivity of $\rho$.}, we obtain the following Fokker--Planck equation:
\al{
\partial_t {}&  \rho =- \partial_{n_1} \l[ \l(\kappa_1 - \kappa_2 n_1^2\r) \rho \r]- \partial_{n_2}\l[ \l(  \frac{\kappa_2}{2} n_1^2 - \kappa_3 n_2 \r) \rho \r] \nonumber \\
{}& + \frac{1}{2\Omega} \l\{ \partial_{n_1}^2\l[ \l(\kappa_1 + \kappa_2 n_1^2\r) \rho \r] + \partial_{n_2}^2\l[ \l(\frac{\kappa_2}{2} n_1^2 + \kappa_3 n_2 \r) \rho \r] \r. \nonumber \\
{}& + \l. \vphantom{\l[\l(\frac{\kappa_2}{2}\r)\r]} 2 \partial_{n_1}\partial_{n_2}\l[ \l( \kappa_2 n_1^2 \r) \rho \r] \r\}. 
}{eq:FP}
The solution to Eq.~\eqref{eq:FP} approximates the stochastic dynamics of the system of interest (Eq.~\eqref{eq:toyModel}) when $\Omega$ is large but finite (i.e., $\Omega \gg 1$). 

It is worth pointing out that in the continuum limit $\Omega \rightarrow \infty$, the Fokker--Planck equation \eqref{eq:FP} reduces to a Liouville equation,
\eq{
\partial_t \rho =- \partial_{n_1} \l[ \l(\kappa_1 - \kappa_2 n_1^2\r) \rho \r]- \partial_{n_2}\l[ \l(  \frac{\kappa_2 n_1^2}{2}  - \kappa_3 n_2 \r) \rho \r].
}{eq:Liouville}
It can be shown that when the initial condition of the probability density function is a Dirac $\delta$-distribution, $\rho\l(x,y,t=0\r)=\delta(x-x_0) \delta\l(y-y_0\r)$, the temporal evolution of the probability density $\rho\l(x,y,t\r)$ remains as a $\delta$-distribution and the peak of the distribution corresponds to the solution of Eqs.~\eqref{eq:rateEq} for any given time $t>0$. Thus, Eq.~\eqref{eq:Liouville} is a slightly generalized version of the equations for mass-action kinetics (Eqs.~\eqref{eq:rateEq}). Eq.~\eqref{eq:Liouville} captures the same information as Eqs.~\eqref{eq:rateEq}, but Eq.~\eqref{eq:Liouville} permits a probabilistic distribution as an initial condition.

When $\Omega$ is large but finite, as is well-known, the dynamics described by a Fokker-Planck equation such as Eq.~\eqref{eq:FP} are largely determined by the equation's advection or drift terms, i.e., the terms in Eq.~\eqref{eq:FP} that are preserved in the Liouville equation (Eq.~\eqref{eq:Liouville}). However, additional terms, i.e., the terms with the operators $\partial_{n_1}^2$, $\partial_{n_2}^2$ and $\partial_{n_1}\partial_{n_2}$, are present in Eq.~\eqref{eq:FP} and these terms introduce diffusion in the state space $\l(n_1,n_2\r)$. It is these terms that characterize density fluctuations that arise from stochastic and discrete reaction events. Because the scale of diffusion is of order $\mathcal{O}\l(1/\Omega\r)$ (i.e., the variance), sample paths $\l(n_1\l(t\r), n_2\l(t\r)\r)$ exhibit fluctuations of order $\mathcal{O}\l(1/\Omega^{1/2}\r)$ (i.e., the standard deviation).

Now let us consider the scaled system, i.e., let us replace $\Omega$ with $\lambda\Omega$ in the above analysis. If $\lambda\Omega$ remains much larger than 1, and we perform the Kramers--Moyal expansion of the CME for the scaled system, we obtain 
\al{
 {}&  \partial_t \rho =- \partial_{n_1} \l[ \l(\kappa_1 - \kappa_2 n_1^2\r) \rho \r]- \partial_{n_2}\l[ \l(  \frac{\kappa_2}{2} n_1^2 - \kappa_3 n_2 \r) \rho \r] \nonumber \\
{}& + \frac{1}{2\Omega\lambda} \l\{ \partial_{n_1}^2\l[ \l(\kappa_1 + \kappa_2 n_1^2\r) \rho \r] + \partial_{n_2}^2\l[ \l(\frac{\kappa_2}{2} n_1^2 + \kappa_3 n_2 \r) \rho \r] \r. \nonumber \\
{}& + \l. \vphantom{\l[\l(\frac{\kappa_2}{2}\r)\r]} 2 \partial_{n_1}\partial_{n_2}\l[ \l( \kappa_2 n_1^2 \r) \rho \r] \r\}. 
}{eq:scaledFP}
This equation is nearly the same as Eq.~\eqref{eq:FP}. The only difference is the factor multiplying the diffusive terms. The factor is $1/(2\Omega)$ in Eq.~\eqref{eq:FP} and $1/(2\Omega\lambda)$ in Eq.~\eqref{eq:scaledFP}. This result generalizes to any chemical reaction system \red{with large species populations (i.e., $N_i \gg 1$ for all $i$)}. 

The significance of our comparison of Eqs.~\eqref{eq:FP}, \eqref{eq:Liouville}, and \eqref{eq:scaledFP} is threefold. First, it is known that first moments are dominated by drift (i.e., the terms with the operators $\partial_{n_1}$ and $\partial_{n_2}$) when $\Omega \gg 1$, and drift is identical in Eqs.~\eqref{eq:FP}, \eqref{eq:Liouville}, and \eqref{eq:scaledFP}. Thus, we can expect simulations of scaled and unscaled systems to yield comparable estimates of first moments when $\Omega$ and $\lambda\Omega$ are large enough such that effects of multiplicative noise can be ignored. Second, we can expect scaling to yield overestimates of second moments, because diffusivity (i.e., variance) is enhanced by a factor of $1/\lambda$ in the scaled system relative to that in the unscaled system. Third, given that standard error of the mean (SEM) is calculated as $\sigma/\sqrt{\mathcal{N}}$, where $\sigma$ is the sample standard deviation and $\mathcal{N}$ is the number of samples, and given that we can expect the sample variance $\sigma^2$ to be amplified by a factor of $1/\lambda$ in estimates of first moments based on scaled simulations, the SEM for each first moment estimate is amplified by a factor of $1/\sqrt{\lambda}$ when using the same number of sample paths from scaled simulations as for estimates based on unscaled simulations. This loss of precision can be overcome by increasing the number of sample paths used. However, the number of sample paths must be increased by a factor of $1/\lambda$ to match the precision of estimates from unscaled simulations. This factor exactly matches the efficiency gain of a single scaled simulation. Thus, scaling yields acceleration \emph{only} when a loss of precision in first moment estimates is permitted.

\subsection{Limitation by small but critical populations}

The Kramers--Moyal expansion of the scaled system is only valid when $\mathcal{O} \l(\lambda N_i\r)\gg 1$ for each $i$ and $\dd n \ll 1$. In the context of scaling, these conditions are satisfied only when $\lambda \gg 1/\min_{1\le i \le M} \l\{N_i\r\}$. When $\lambda$ is too small, one or more populations become discrete. Discrete populations have distinct dynamics, which are not captured by a Fokker--Planck equation  \cite{lin2016gene,lin2016bursting,lin2018efficient}. Thus, the analysis of the previous section eventually becomes invalid and errors are introduced as $\lambda$ decreases and becomes too small. For example, for the system of Eq.~\eqref{eq:toyModel}, $\lambda = 3 \times 10^{-4}$ is evidently too small, as indicated by the errors in the estimates of the first moments that can be seen in Fig.~\ref{fig:dimerization}(e) and (g). These errors arise because the scaled system is no longer representative of the unscaled system.

Scaling a system so that the smallest population vanishes (i.e., becomes 0) or reduces to order 1 might be acceptable in practice but requires careful consideration. On one hand, the population of concern may be of little significance for system behavior, and the erroneous behavior predicted for it can be safely ignored. On the other hand, if the population is critical for system behavior, then scaling will produce seriously misleading results. 

Specialized methods are available for coping with discrete populations that influence system dynamics so long as these populations are always discrete and well-separated from other populations \cite{bokes2013transcriptional,lin2016gene,lin2016bursting,lin2018efficient}. However, for many systems, populations are not only distributed over a spectrum of scales but also dynamically changing scale, as illustrated in Fig.~\ref{fig:models}. We suspect that this behavior is generic, at least for biochemical systems, which limits the applicability of static and homogeneous scaling. Below, we present a new scaling approach that overcomes this limitation.

\renewcommand{\algorithmicrequire}{}
\renewcommand{\algorithmicensure}{\begin{footnotesize}\rule[10pt]{\textwidth}{0.3pt}\end{footnotesize}}

\begin{figure*}
\begin{minipage}{\linewidth}
\begin{algorithm}[H]
\caption{Partial scaling algorithm (PSA).} \label{alg:partialScaling}
   \begin{algorithmic}[1]
\Require This algorithm produces a sample path of the stochastic dynamics of a chemical reaction network. The following inputs are required to execute the algorithm. (1) A chemical reaction network consisting of $R$ reactions among $M$ chemical species. Reactions are indexed by $r\in \l\{1\ldots R\r\}$ and chemical species are indexed by $s\in S \equiv \l\{1\ldots M\r\}$. (2) The number of reactant and product species participating in each reaction $r$,  $\mu_r \in \mathbb{N}$ and $\nu_r\in \mathbb{N}$, respectively\footnote{For simplicity, we assume $\mu_r,\nu_r>0$, but the algorithm can be trivially generalized to account for $\mu_r=0$ (e.g., creation) or $\nu_r=0$ (e.g., annihilation).}. (3) The set of indices of reactant and product species participating in each reaction $r$, $\l\{\alpha_{r,k} \r\}_{k=1}^{\mu_r}$ and $\l\{\beta_{r,k} \r\}_{k=1}^{\nu_r}$, where each $\alpha_{r,k}, \beta_{r,k} \in S$. (4) The propensity function of each reaction $r$, $f_r\l(\mathbf{N}\r) \equiv h_r(\mathbf{N})\bar{\kappa}_r$. (5) The stoichiometric coefficient $\xi_{r,s}$ for each species $s$ participating in each reaction $r$. We take stoichiometric coefficients to be negative for reactants and positive for products. (6) The initial condition, $\l\{N_s\l(t=0\r)\r\}_{s=1}^{M}$. (7) A list of $K$ discrete times $0 \le t_1 \le t_2 \le \ldots t_K$ at which system state should be reported. The last input is a critical population scale $N_c$.
\newcommand{\algrule}[1][.2pt]{\par\vskip.3\baselineskip\hrule height #1\par\vskip.5\baselineskip}
\algrule
\State{$t \gets 0$} \Comment{Initialize time $t$}
\For{$s$ in $\l\{1,\ldots M\r\}$} 
\State{$N_s \gets N_s(t=0)$} \Comment{Initialize each population $N_s$}
\EndFor
\For{$r$ in $\l\{1,\ldots R\r\}$} 
\State{$S_r \gets \l\{\alpha_{r,1}, \ldots, \alpha_{r,\mu_r}\r\} \cup \l\{\beta_{r,1}, \ldots, \beta_{r,\nu_r}\r\}$} \Comment{Create a list of chemical species that participate in each reaction $r$}
\EndFor
\For{$i$ in $\l\{1,\ldots K\r\}$} \Comment{Start continuous-time Markov chain (CTMC)} 
\While{$t< t_i$} 
\For{$r$ in $\l\{1,\ldots R\r\}$} 
\State $\lambda_r \gets 1/\max\l\{1, \l\lfloor\frac{\min_{s\in S_r} \l\{N_s(t)\r\}}{N_c}\r\rfloor \r\}$ \Comment{Calculate the scaling factor $\lambda_r$ for each reaction $r$}
\State $\kappa_r \gets \lambda_r f_r \l(\mathbf{N}\r)$ \Comment{Calculate the scaled rate of reaction for each reaction $r$} 
\EndFor
\State $\kappa \gets \sum_{r=1}^R \kappa_r$ 	\Comment{Calculate the scaled overall rate of reaction}
\State $\Delta t \gets \text{Exp}\l(\kappa\r)$		\Comment{Generate a random waiting time to the next reaction}
\If{$t + \Delta t < t_i $} \Comment{Check if the advanced time is ahead of the next report time}
\State $t \gets t+\Delta t$ \Comment{Increment time}
\State $\phi \gets \kappa \times \text{Unif}(0,1)$ \Comment{Inverse sampling to select the next reaction $r$}
\State $r \gets 1$ 
\While{$\sum_{j=1}^{r}\kappa_j<\phi$} 
\State $r \gets r+1$ 
\EndWhile 
\For{$s$ in $S_r$}
\State $N_s \gets N_s + (1/\lambda_r) \times \xi_{r,s}$ \Comment{Update the populations in $S_r$ (using scaled stoichiometric coefficients)}
\EndFor
\Else
\State $t\gets t_i$ \Comment{Advance time to the next report time without updating the system state}
\State Export state vector $\mathbf{N}(t)$ \Comment{Report system state (all populations)}
\EndIf
\EndWhile
\EndFor \Comment{Perform next step in CTMC until reports have been made for all specified report times}
   \end{algorithmic}
\end{algorithm}
\end{minipage}
\end{figure*}

\section{Adaptive and heterogeneous scaling}

There are two barriers to applying  static and homogeneous scaling. First, population scales may be scattered across a broad spectrum (Fig.~\ref{fig:models}). Second, populations may evolve dynamically across multiple scales (Fig.~\ref{fig:models}). To address these problems, we propose a new scaling approach, which we term \emph{partial scaling}, that, in contrast with standard scaling, is adaptive and heterogeneous, as we will see. Both of these features of partial scaling arise from tying scaling to population sizes, on the fly. Scaling is heterogeneous because population sizes are heterogeneous, and scaling is adaptive because populations are dynamic. A particular, intentionally simple implementation of partial scaling is outlined as pseudocode in Algorithm \ref{alg:partialScaling}. We call this algorithm the partial scaling algorithm (PSA).

With partial scaling, there is no longer a global scaling factor $\lambda$, as in standard scaling. Instead, at any given time $t$, there is a scaling factor $\lambda_r(t)$ for each reaction $r=1,\ldots,R$. These scaling factors are not set in advance of a simulation, nor statically. Rather, they are assigned values according to a dynamic update schedule (e.g., a schedule synchronized with time updates). This update schedule ensures that
\eq{
\lambda_r\l(t\r) = \frac{1}{\max \l\{1, \l\lfloor \frac{N_{\min}^r\l(t\r)}{N_c}\r\rfloor \r\}},
}{eq:update_formula}
where $N_c$ is a threshold or critical population---this quantity is the \emph{one} parameter of the method---and $N_{\min}^r\l(t\r)$ is the smallest population among those of the reactants and products of reaction $r$ at time $t$. It should be noted that the brackets around the ratio $N^r_{\min}/N_c$ in Eq.~\eqref{eq:update_formula} denote the floor function and, furthermore, that there is no scaling (i.e., $\lambda_r=1$) for reaction $r$ if any reactant or product population is smaller than $N_c$.

The scaling factors $\{ \lambda_r(t) \}_{r=1}^R$ defined by Eq.~\eqref{eq:update_formula} are used to modify an SSA, such as Gillespie's direct method \cite{gillespie1977exact}, as follows. In the calculation of stochastic transition rates (e.g., as for Eqs.~\eqref{eq:sto} as we have discussed), the rate for each reaction $r$ is calculated as usual but then scaled (down) by a factor $\lambda_r$. When reaction $r$ with rate scaled by $\lambda_r<1$ is selected to fire, to compensate for its reduced rate of firing, its stoichiometric coefficients are each temporarily scaled (up) by the factor $1/\lambda_r$. The amplified stoichiometric coefficients are then used to update the system state $\mathbf{N}$. In other words, $\mathbf{N}$ is modified in accordance with scaled population changes triggered by the selected reaction event. It should be noted that $1/\lambda_r \in \mathbb{N}$ by construction (Eq.~\eqref{eq:update_formula}). Thus, scaled stoichiometric coefficients, like unscaled coefficients, are natural numbers. A specific example of how the scaling factors $\{ \lambda_r(t) \}_{r=1}^R$ are used to modify an SSA is provided in Algorithm \ref{alg:partialScaling}. It should be noted that Algorithm \ref{alg:partialScaling} reduces to Gillespie's direct method if we statically assign each $\lambda_r$ a value of 1. Furthermore, standard scaling is achieved by replacing each reaction-specific adaptive scaling factor $\lambda_r$ in Algorithm \ref{alg:partialScaling} with a universal static scaling factor $\lambda$.

With partial scaling, population sizes are not directly scaled but the populations of reactants and products participating in any reaction $r$ for which $\lambda_r<1$ are effectively $\mathcal{O}(N_c)$ or larger. Thus, a choice of $N_c\gg 1$ guarantees that stochastic dynamics do not become inappropriately bursty. Furthermore, any critical species $X_i$ with a small population $N_i<N_c$ will never be scaled out of the system. In other words, $N_i$ is never effectively scaled such that the population vanishes because $\lambda_r=1$ for any reaction having a reactant or product population less than $N_c$ (Eq.~\eqref{eq:update_formula}).

\subsection{Accuracy and precision of partial scaling}

As we will see in this section, provided that $N_c \gg 1$, partial scaling preserves means but not variances, similar to standard scaling.

Let us consider a time interval $\Delta t \ll ( \sum_{r=1}^R h_r \bar{\kappa}_r)^{-1}$, i.e., an interval that is much shorter than the expected waiting time to the next reaction event. Furthermore, consider an arbitrary species $X_i$, $i\in\{1,\ldots,M\}$, which participates, as a reactant or product, in a set of reactions with indices $I\subseteq\l\{1,\ldots, R\r\}$. We will use $I^{\prime} \subseteq I$ to denote the subset of these indices corresponding to reactions associated with a scaling factor less than 1. Thus, for each $r \in I^{\prime}$, $\lambda_r<1$, and for each $r \in I \backslash I^{\prime}$, $\lambda_r=1$. For an unscaled simulation, we will use $\delta N_{i,r}$ to denote the change of population $N_i$ brought about by a reaction $r \in I$ occurring within the time interval $\Delta t$. Given that multiple reaction events can occur within the time window $\Delta t$ (albeit with low probability, by construction), the total change in $N_i$ over this time window, which we will denote as $\delta N_i$, is given by
\eq{
\delta N_i = \sum_{r\in I} \delta N_{i,r}.
}{eq:unscaledN}
Similarly, for a (partially) scaled simulation, the total population change is given by
\eq{
\delta N_i^\text{scaled} = \sum_{r\in I^{\prime}} \frac{\delta N_{i,r}}{\lambda_r} + \sum_{r\in I \backslash I^{\prime}} \delta N_{i,r}.
}{eq:scaledN}
This expression is the same as that given above except for the introduction of scaling factors for stoichiometric coefficients. 

The population changes considered above (i.e., the $\delta N_i$ and $\delta N_{i,r}$ terms in Eqs.~\ref{eq:unscaledN} and \ref{eq:scaledN}) are random variables. Because $\Delta t$ is small (relative to the expected waiting time to the next reaction), we can take each of these random variables to be drawn from a Poisson distribution. Thus, 
\subeq{
\delta N_{i,r} \sim{}& \text{Poisson}\l( h_r \bar{\kappa}_r \Delta t\r), \\
\delta N_{i,r}^{\star} \sim{}& \lambda_r^{-1} \text{Poisson}\l( \lambda_r h_r \bar{\kappa}_r \Delta t\r), 
}{}
where $\delta N_{i,r}^{\star} \equiv \delta N_{i,r}/\lambda_r$ denotes a population change triggered by firing of a reaction $r$ with a scaled rate $\lambda_r h_r \bar{\kappa}_r$.

\red{For constants $c$, $\mu>0$, recall that $\E{c \cdot \text{Poisson}\l(\mu \r)} = c \mu$ and $\var{c \cdot \text{Poisson}\l(\mu \r)} =c^2 \mu$.} Thus, for chemical species $X_i$ and reaction $r \in I^{\prime}$,\red{
\subeq{
\E{ \delta N_{i,r}^{\star}} ={}& \frac{\lambda_r h_r \bar{\kappa}_r \Delta t}{\lambda_r} = h_r \bar{\kappa}_r \Delta t = \E{\delta N_{i,r}}, \label{eq:13a}\\
\var{ \delta N_{i,r}^{\star}} ={}& \frac{ \lambda_r h_r \bar{\kappa}_r \Delta t}{\lambda_r^2} > h_r \bar{\kappa}_r \Delta t = \var{ \delta N_{i,r}}.\label{eq:13b}
}{}
} It should now be clear that partial scaling yields unbiased estimates of first moments and overestimates of second moments \red{(in a time interval $\Delta t \ll 1$)}, as is also the case for standard scaling. However, the bias in estimates of second moments, overall, is less with partial scaling than with standard scaling, because the processes in a (partially) scaled simulation that are unaffected by scaling have the same statistics as the processes in an unscaled simulation.

\red{In partial scaling, each time a reaction $r$ (with scaled rate $\lambda_r h_r \bar{\kappa}_r$) is fired, a \emph{deterministic multiplier}, $1/\lambda_r$, is applied to the reaction's stochiometric coefficients for the purpose of appropriately updating the system state (Algorithm~\ref{alg:partialScaling}, line 24). As indicated by the analysis above, this multiplier is essential for obtaining unbiased estimates of first moments. Estimates of first moments would still be unbiased if the multiplier were instead chosen randomly with an expected value of $1/\lambda_r$; however, estimates of variance would be more biased for a random multiplier than for a deterministic multiplier\footnote{\red{Let us use $XY$ to denote the population change of a reaction event (in a partially scaled simulation), where $X$ is a possibly random multiplier and $Y\sim \text{Poisson}\l(\lambda_r \bar{\kappa}_r h_r \Delta t \r)$. If the multiplier is distributed such that $\mathbb{E}\l[X\r] = 1/\lambda_r$ and $X$ and $Y$ are independent, $\E{XY}=h_r\bar{\kappa}_r \Delta t$ (as desired for preservation of first moments) and  $\var{XY}= \var{X} \lambda_r h_r \bar{\kappa}_r \Delta t \l(1 + \lambda_r h_r \bar{\kappa}_r \Delta t \r) + h_r \bar{\kappa}_r \Delta t /\lambda_r$. Because all terms in this expression are positive, it follows that $\var{XY}$ is least when $\var{X}=0$, i.e., when $X$ is a deterministic variable. Recall that for two independently distributed random variables $X$ and $Y$, $\E{XY} =\E{X}\E{Y}$ and $\var{XY} = \var{X}\, \var{Y} + \var{X} \, \mathbb{E}^2\l[Y\r] + \mathbb{E}^2\l[X\r]\,\var{Y}$.}}.}

\red{One may wonder if partial scaling yields unbiased first-moment estimates in the short term but not in the long run. We expect that the answer is model dependent. If a model permits noise-induced phenomena involving chemical species participating in reactions with scaled rates, then estimates may become biased because of demographic fluctuations. In other words, systematic errors may arise with partial scaling from behavior that depends on second or higher moments. Below, in part to address this issue, we turn our attention to numerical experiments, including simulations for a model that permits bistable stochastic switching,  the TCR model \cite{lipniacki2008stochastic}. As we will see, partial scaling yields unbiased estimates of first moments (and also higher-order moments), even for this model.}

\begin{figure*}[t]
\begin{center}
\includegraphics[width=1.0\textwidth]{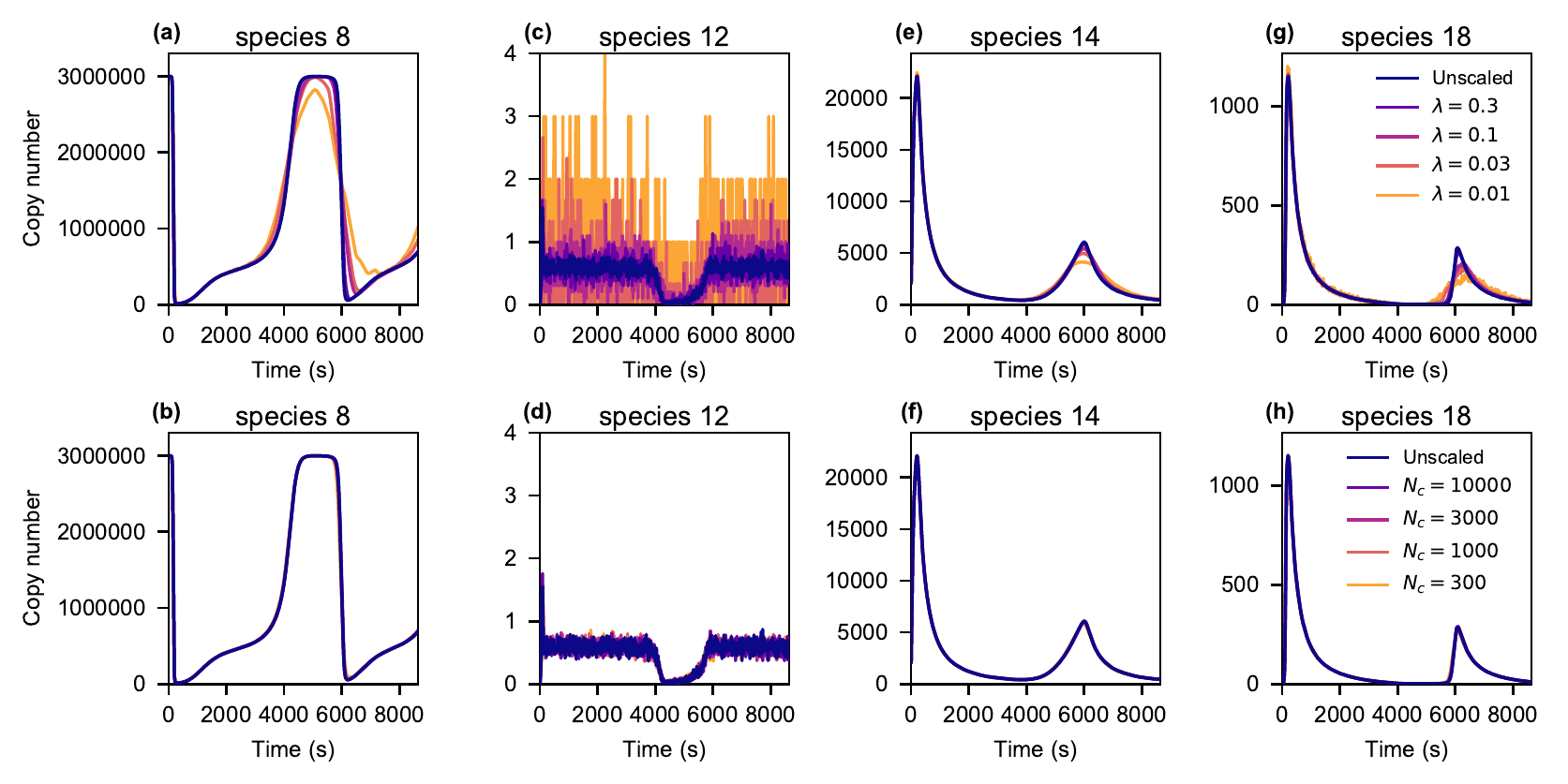}
\end{center}
\caption{Comparison of the standard scaling method (a,c,e,g) with the partial scaling method (b,d,f,h) for simulations based on a model for  ERK activation \cite{kochanczyk2017relaxation}. Shown here are the mean populations for four selected chemical species, labeled 8, 12, 14, and 18. These labels are those produced by BioNetGen; the species are defined in the NET file produced as output when BioNetGen processes the BNGL file defining the ERK activation model\cite{ERK_BNGL}. Means are based on 500 sample paths. Results from standard scaling begin to deviate from those obtained for the unscaled system via exact simulation when the scaling factor is still fairly large ($\lambda=0.1$). The partial scaling method yields reasonable estimates for both first and second moments, even with very aggressive scaling ($N_c=300$). Compare 300 to the largest population size in Fig.~\ref{fig:models}(a), which is on the order of $10^6$.}
\label{fig:ERK}
\end{figure*}

\section{Comparison of scaling methods}

To facilitate benchmarking of partial scaling, we implemented a variation of Gillespie's direct method, an additional variation that incorporates standard scaling, and a variation of PSA (Algorithm~\ref{alg:partialScaling}). Importantly, we implemented these methods so as to eliminate all unnecessary implementation differences. We used our code to perform stochastic simulations based on three published models \cite{kochanczyk2017relaxation,rubenstein2007dynamics,lipniacki2008stochastic} with the goal of evaluating the relative efficiency of each of the three methods (no scaling, standard scaling and partial scaling) and the relative accuracy of standard and partial scaling. In simulations, we used a model-specific pre-generated dependency graph to aid in updates of propensity functions but we did not use sorting of the propensity functions to accelerate simulations as in the Gibson-Bruck method \cite{gibson2000efficient}. In our simulations, we considered different settings for the algorithmic parameters $\lambda$ (the static scaling factor in standard scaling) and $N_c$ (the critical population size used in partial scaling).

The models that we considered, all for biological systems, were chosen because each was originally analyzed using exact stochastic simulation and each challenges the application of standard scaling. The first of the three models characterizes activation of ERK \cite{kochanczyk2017relaxation}. The dynamics of this model are oscillatory. Populations span multiple orders of magnitude and their scales change dynamically (Fig.~\ref{fig:models}(a)). The second of the three models characterizes prion protein aggregation \cite{rubenstein2007dynamics}, which involves polymerization-like reactions and seeded nucleation of aggregates. As before, the populations considered in this model evolve over multiple scales, and the initial abundance of prion protein is discrete (Fig.~\ref{fig:models}(b)). The third of the three models characterizes TCR signaling \cite{lipniacki2008stochastic}. The behavior characterized by this model includes stochastic bistable switching. In other words, for a particular regime of behavior, which we considered in our simulations (as is evident from inspection of individual sample paths), intrinsic noise is capable of driving the system from one stable fixed point to another (and back). This type of behavior has been intensely studied in the context of genetic regulatory circuits \cite{walczak2005self,roma2005optimal,warren2005chemical,assaf2011determining,strasser2012stability,bokes2013transcriptional,lu2013tristability,lin2016bursting}. For the TCR model, unlike for many models of gene regulation that have been considered, many of the populations are large (i.e., near the continuum limit) (Fig.~\ref{fig:models}(c)).

\begin{figure*}[t]
\begin{center}
\includegraphics[width=1.0\textwidth]{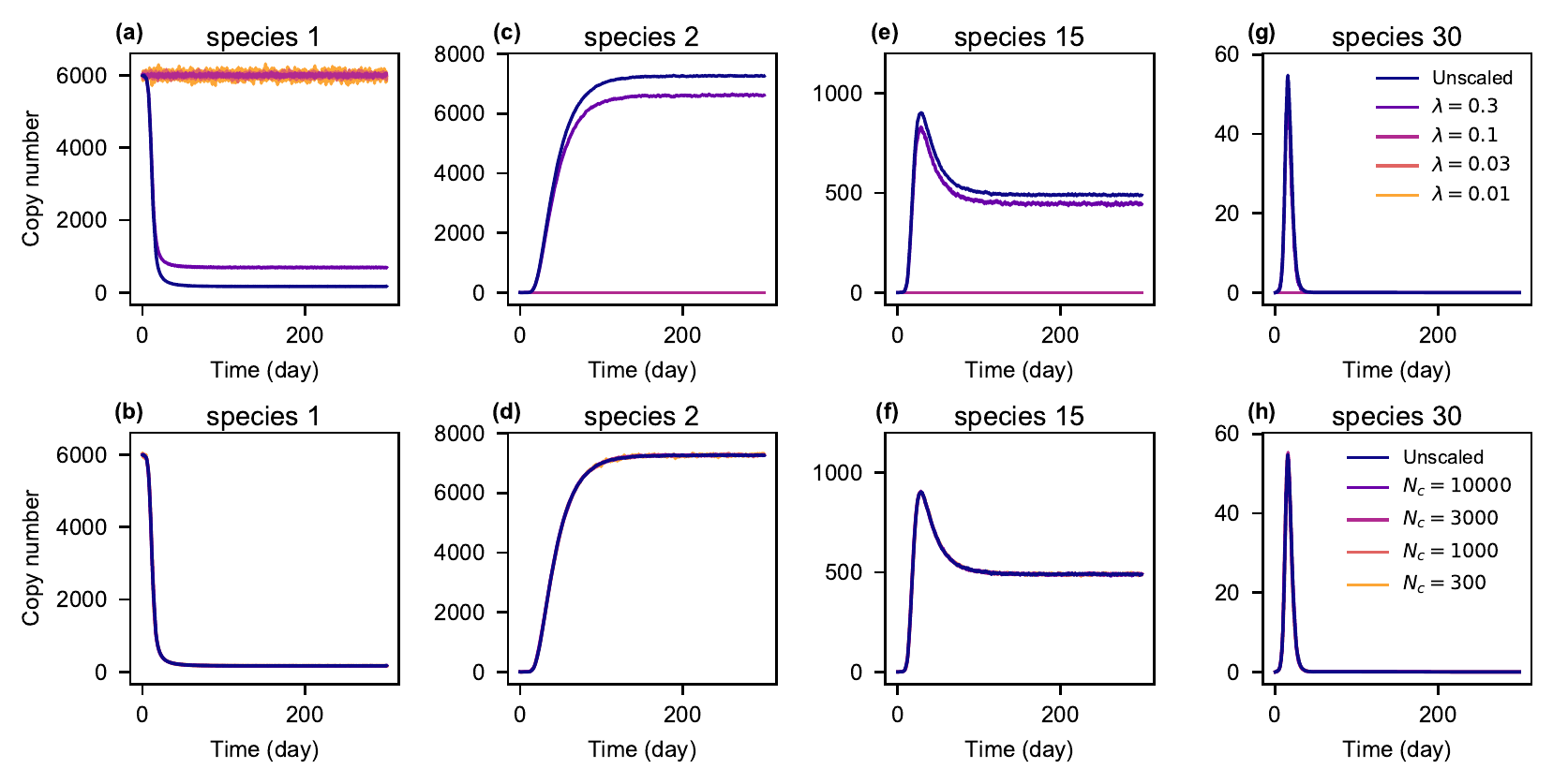}
\end{center}
\caption{Comparison of the standard scaling method (top panels) with the partial scaling method (bottom panels) for simulations based on a model for prion protein aggregation \cite{rubenstein2007dynamics}. Mean abundances of four selected chemical species, labeled 1, 2, 15 and 30, are calculated from 500 sample paths. The species labels are those produced by BioNetGen; the species are defined in the NET file produced as output when BioNetGen processes the BNGL file defining the prion protein aggregation model\cite{Prion_BNGL}. The standard scaling method produces results that deviate from those generated for the unscaled system via exact simulation even when the scaling factor is fairly large ($\lambda=0.3$). This finding is explained by the discreteness of the species seeding prion protein aggregation. The partial scaling method produces reasonable estimates of both first and second moments even with aggressive scaling ($N_c =300$). Compare 300 with the largest population size in Fig.~\ref{fig:models}(b), which is on the order of $10^3$ to $10^4$.}
\label{fig:prion}
\end{figure*}

\begin{figure*}[t]
\begin{center}
\includegraphics[width=1.0\textwidth]{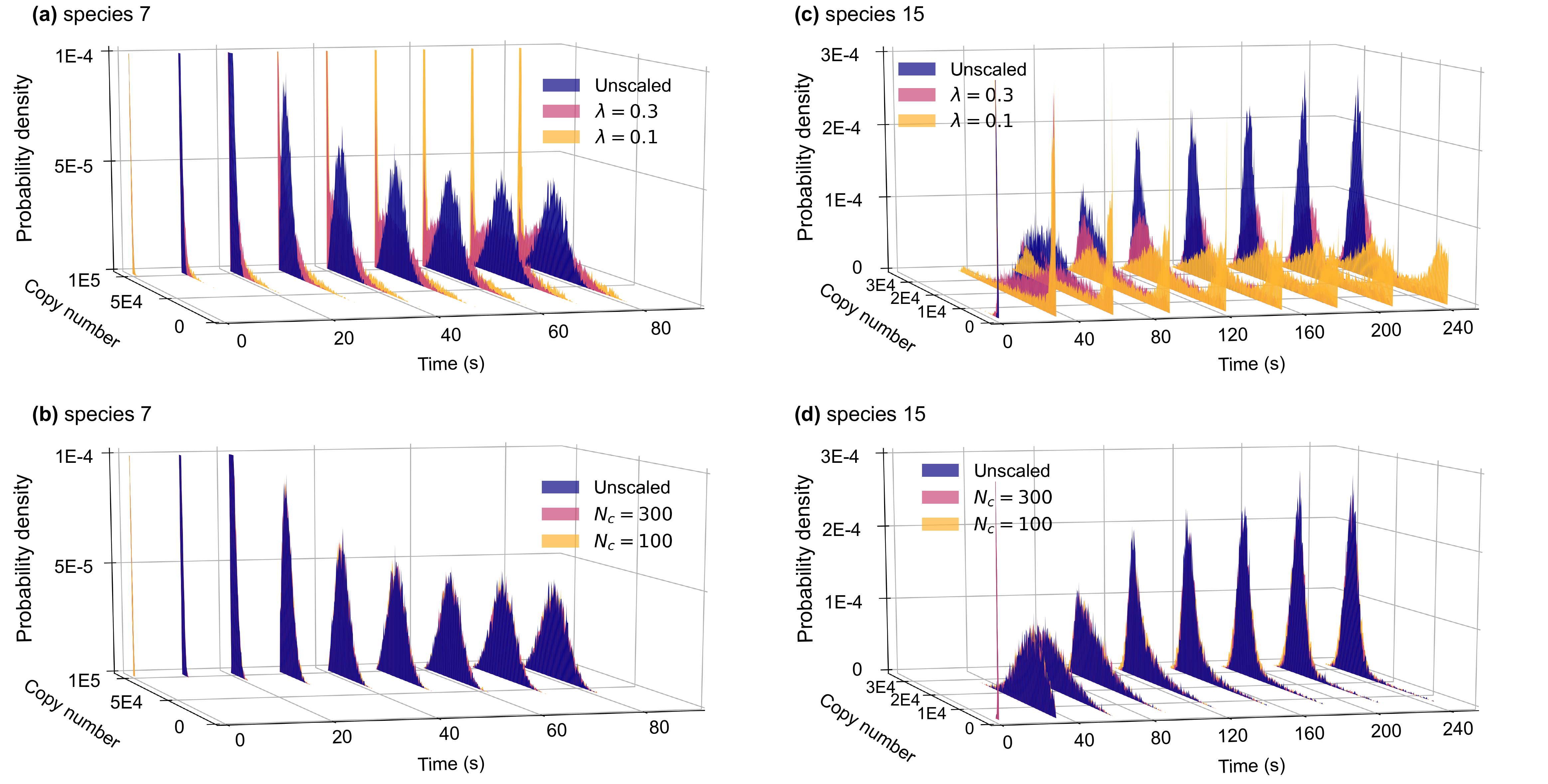}
\end{center}
\caption{Comparison of the standard scaling method (top panels) with the partial scaling method (bottom panels) for simulations based on a model for TCR signaling \cite{lipniacki2008stochastic}. Shown here are the time-dependent marginal probability distributions for two selected chemical species considered in the model, which are labeled 7 and 15. These labels are those produced by BioNetGen; the species are defined in the NET file produced as output when BioNetGen processes the BNGL file defining the TCR signaling model\cite{TCR_BNGL}. The marginal probability distributions have been constructed on the basis of $10^4$ sample paths. The standard scaling method produces results that deviate from those for the unscaled system even when the scaling factor is fairly large ($\lambda=0.3$). In contrast, the partial scaling method generates distributions that are comparable to those calculated on the basis of exact simulations even with very aggressive scaling ($N_c=100$). Compare 100 with the largest population size in Fig.~\ref{fig:models}(c), which is on the order of $10^5$ to $10^6$.}
\label{fig:TCRdistribution}
\end{figure*}

\begin{figure*}[t]
\begin{center}
\includegraphics[width=0.7\textwidth]{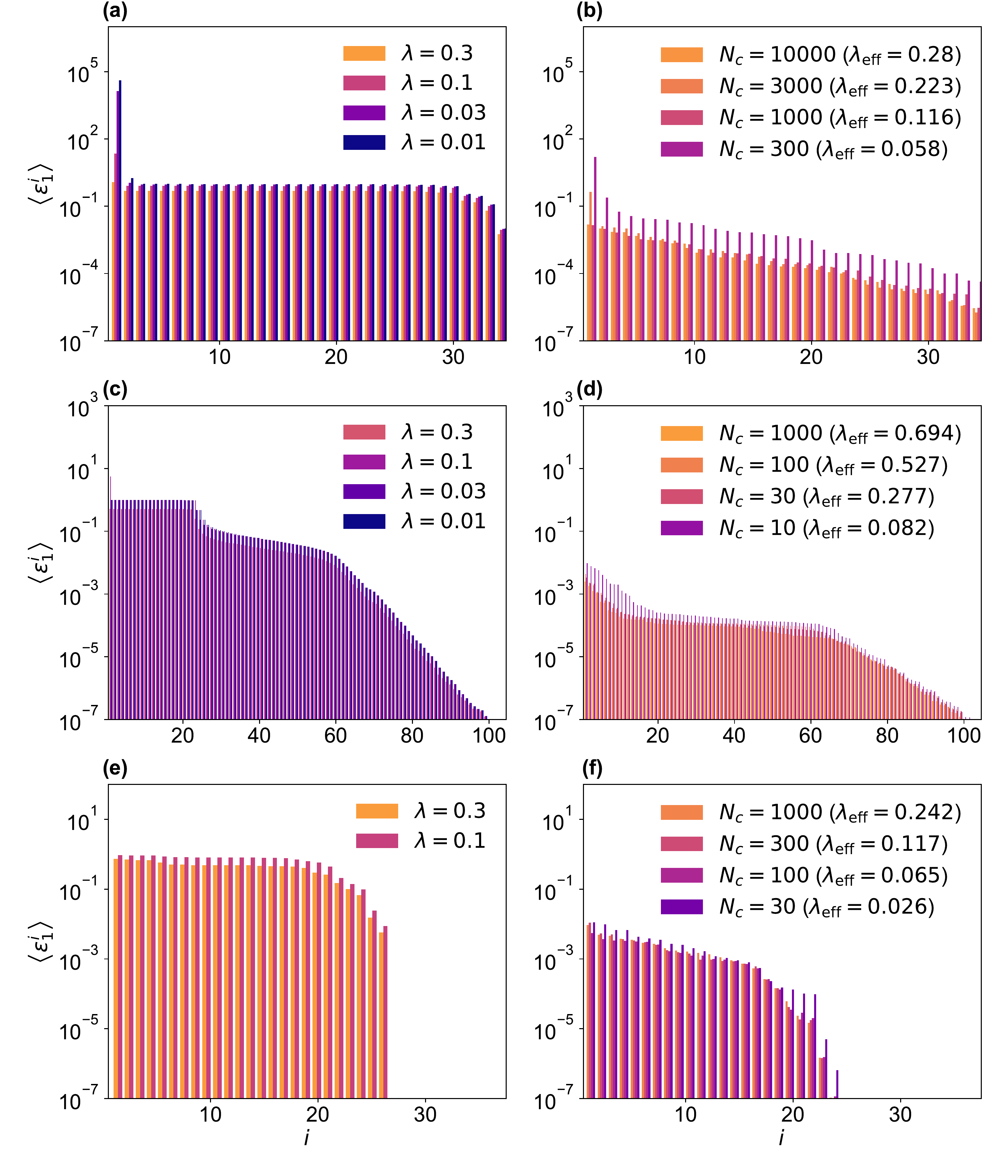}
\end{center}
\caption{\red{Summary of numerical experiments aimed at quantifying the relative accuracy of partial scaling. First-moment errors introduced by standard scaling (left) and partial scaling (right) in simulations of the ERK activation model (top), prion protein aggregation model (middle), and TCR signaling model (bottom). As described in the Appendix, temporally averaged errors were calculated for the ensemble averaged trajectories of individual species. Errors are plotted in each panel in order, from greatest (left) to least (right). The species indices are hidden; the scale of each horizontal axis matches the number of species in the corresponding model. It should be noted that each vertical axis is logarithmic, with the scale varying from model to model. The values of $\lambda_{\rm eff}$ indicated in the legends at right indicate aggressiveness of scaling (see Appendix); these values can be compared to the values of $\lambda$ indicated in the legends at left. Overall first- and second-moment errors for the numerical experiments considered here are reported in Table~\ref{table:accuracyEfficiency}.}}
\label{fig:errors}
\end{figure*}

In Figs.~\ref{fig:ERK} and \ref{fig:prion}, we plot predicted time courses for selected chemical species considered in the ERK model\cite{kochanczyk2017relaxation} and the prion model\cite{rubenstein2007dynamics}, respectively. Each point in each time course corresponds to the mean population for the indicated chemical species estimated on the basis of 500 sample paths. \red{(This number of sample paths was deemed sufficient for accurate estimation of statistical quantities of interest.)} In each figure, the top row of panels shows results of calculations based on either exact simulation or standard scaling (for different values of $\lambda$), as indicated in the legend at right. Similarly, in each figure, the bottom row of panels shows results of calculations based on either exact simulation or partial scaling (for different values of $N_c$), as indicated in the legend at right. As can been, standard scaling yields estimates that deviate markedly from those based on exact simulation. In contrast, even with aggressive scaling ($N_c=300$), partial scaling yields first-moment estimates that are barely distinguishable from those based on exact simulation.

In Fig.~\ref{fig:TCRdistribution}, we plot time-dependent marginal probability distributions (estimated on the basis of $10^4$ sample paths) for two selected chemical species considered in the TCR model\cite{lipniacki2008stochastic}. \red{(We considered $10^4$ vs. 500 sample paths because here we are estimating distributions vs. properties of distributions.)}  In this figure, we focus on the initial transient shown in Fig.~\ref{fig:models}(c). The dynamics of this model are inherently stochastic and poorly represented by first moments. As can be seen, standard scaling yields marginalized probability distributions that differ markedly from those obtained from exact simulations (top row), whereas partial scaling yields approximate results that are very close to the exact results (bottom row).

Additional accuracy results, based on error measures introduced in the Appendix, and timing results for the three benchmark problems \red{are summarized in Fig.~\ref{fig:errors} and} Table~\ref{table:accuracyEfficiency}. At least for these problems, partial scaling allows for more significant speed ups than standard scaling with the introduction of much less error in the estimates of first and second moments. 

\begin{figure*}[t]
\begin{center}
\includegraphics[width=1.0\textwidth]{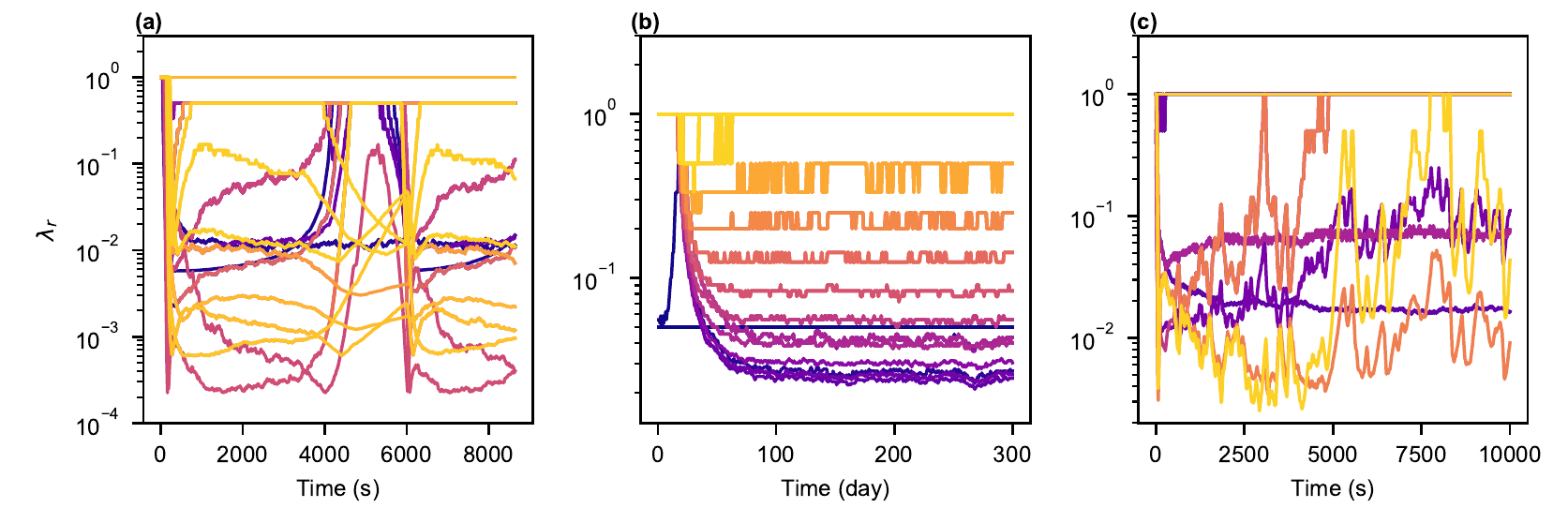}
\end{center}
\caption{Values of $\lambda_r (t)$ in a single stochastic simulation performed using the partial scaling method for every reaction $r\in \l\{1\ldots R\r\}$ in the (a) ERK activation model \cite{kochanczyk2017relaxation}, (b) prion protein aggregation model \cite{rubenstein2007dynamics}, and (c) TCR signaling model \cite{lipniacki2008stochastic}. In each panel, $N_c=300$. As can be seen, scaling is adaptive and heterogeneous.}
\label{fig:lambda-t}
\end{figure*}

In Fig.~\ref{fig:lambda-t}, for one particular setting for $N_c$, we have plotted $\lambda_r$ as a function of time $t$ for every reaction included in each of the three models considered above. These plots show that partial scaling is highly dynamic and multiscale.

\section{Implementation}

We added a partial scaling feature to the SSA implemented in BioNetGen\cite{Harris2016}, an open-source, general-purpose simulation package used by biological modelers. The SSA implemented in BioNetGen is an efficient variation of Gillespie's direct method \cite{Gillespie2007}, which incorporates various ideas for optimizing simulation efficiency, such as on-the-fly generation (vs. pre-generation) of the list of reactions in which (populated) chemical species can participate \cite{Lok2005,Faeder2005}. BioNetGen is designed for compatibility with models defined using BNGL \cite{Faeder2009}, a language for specifying deterministic and stochastic models for well-mixed (bio)chemical reaction kinetics and for specifying simulations based on such models. BioNetGen also supports models defined using SBML\cite{Hucka2018}, such as those available in BioModels Database\cite{Chelliah2013}. To invoke partial scaling using BioNetGen's \texttt{simulate} command, the \texttt{method} argument should be set to \texttt{ssa} \red{or \texttt{psa}} and a new \red{\texttt{poplevel}} argument, which is used only with partial scaling, should be assigned a non-negative integer value. For accuracy, care should be taken to assign a value much larger than 1. The setting for \red{\texttt{poplevel}} determines the value of $N_c$ and thereby the aggressiveness of scaling. Examples of usage are provided in BNGL files that we have made available online \cite{ERK_BNGL,Prion_BNGL,TCR_BNGL}. These files define the ERK activation\cite{kochanczyk2017relaxation}, prion protein aggregation\cite{rubenstein2007dynamics}, and TCR signaling\cite{lipniacki2008stochastic} models considered in Figs.~\ref{fig:ERK}--\ref{fig:TCRdistribution}. It should be noted that our implementation of partial scaling in BioNetGen \red{(version 2.5 or higher)} is a so-called generate-first method, meaning that it requires an enumeration of chemical species and the individual reactions in which these species are able to participate. In contrast, so-called network-free methods do not require an explicit enumeration of chemical species or reactions \cite{Suderman2018}.

\begin{table*}[t]
\begin{tabular}{@{}lllll@{}}
Model & Scaling method & CPU time & \red{ $\l\langle \varepsilon_1 \r\rangle$ } & \red{$\l\langle CV_\text{VV} \r\rangle$}  \\
\toprule 
ERK & None (Gillespie SSA)  &  $5.30\times10^2\pm	1.17\times10^1$         &$ 0.00\times 10^{0} $ & $ \red{2.98\times 10^{-2}} $ \\
ERK & Standard ($\lambda=0.3$)  &  $1.57\times10^2\pm	6.10\times10^0$         &$ \red{4.36\times 10^{-1}}$ & $\red{5.39\times 10^{-2}} $ \\
ERK & Standard ($\lambda=0.1$)  &  $5.21\times10^1\pm	2.42\times10^0$         &$ \red{1.31\times 10^{0}}$ & $\red{9.44\times 10^{-2}}$ \\
ERK & Standard ($\lambda=0.03$)  &  $1.52\times10^1\pm	1.10\times10^0$         &$ \red{3.99\times 10^{+2}}$ & $ \red{1.77\times 10^{-1}} $ \\
ERK & Standard ($\lambda=0.01$)  &  $4.63\times10^0\pm	5.12\times10^{-1}$      &$ \red{1.20\times 10^{+3}}$ & $\red{2.79\times 10^{-1}} $ \\
ERK & Partial ($N_c=3.0\times10^4$)  &  $4.83\times10^2\pm	1.01\times10^1$             &$ \red{3.44\times 10^{-3}}$ & $ \red{3.03\times 10^{-2}} $ \\
ERK & Partial ($N_c=1.0\times10^4$)  &  $4.14\times10^2\pm	9.25\times10^0$             &$ \red{1.75\times 10^{-3}}$ & $ \red{3.17\times 10^{-2}} $ \\
ERK & Partial ($N_c=3.0\times10^3$)  &  $2.82\times10^2\pm	8.42\times10^{0}$           &$ \red{1.47\times 10^{-2}}$ & $ \red{3.49\times 10^{-2}} $ \\
ERK & Partial ($N_c=1.0\times10^3$)  &  $9.95\times10^1\pm	4.42\times10^{0}$           &$ \red{1.67\times 10^{-3}}$ & $\red{4.50\times 10^{-2}}$ \\
ERK & Partial ($N_c=3.0\times10^2$)  &  $4.47\times10^1\pm	2.48\times10^{0}$           &$ \red{4.80\times 10^{-1}}$ & $ \red{6.94\times 10^{-2}}$ \\
\hline
Prion & None (Gillespie SSA)  &  $4.39\times10^2\pm	5.24\times10^1$         &$ 0.00\times 10^{0} $ & $ \red{2.03\times 10{-2}} $\\
Prion & Standard ($\lambda=0.3$)  &  $1.25\times10^2\pm1.38\times10^1$           &$ \red{1.22\times 10^{-1}}$ & $\red{3.00\times 10{-1}} $\\
Prion & Standard ($\lambda=0.1$)  &  $2.61\times10^1\pm	7.61\times10^{-1}$      &$ \red{2.87\times 10^{-1}} $ & $ \red{3.92\times 10{-2}} $\\
Prion & Standard ($\lambda=0.03$)  &  $8.19\times10^0\pm	4.90\times10^{-1}$      &$ \red{2.35\times 10^{-1}} $ & $ \red{5.73\times 10{-2}} $\\
Prion & Standard ($\lambda=0.01$)  &  $3.08\times10^0\pm	2.68\times10^{-1}$      &$ \red{2.39\times 10^{-1}} $ & $ \red{5.96\times 10{-2}} $\\
Prion & Partial ($N_c=3.0\times10^3$)  &  $6.29\times10^2\pm	6.51\times10^0$             &$ \red{1.34\times 10^{-4}}$ & $ \red{2.18\times 10{-2}} $\\
Prion & Partial ($N_c=1.0\times10^3$)  &  $4.50\times10^2\pm	5.67\times10^0$             &$ \red{1.33\times 10^{-4}}$ & $ \red{2.79\times 10{-2}} $\\
Prion & Partial ($N_c=3.0\times10^2$)  &  $4.18\times10^2\pm	5.56\times10^{0}$           &$ \red{1.32\times 10^{-4}}$ & $ \red{3.95\times 10{-2}} $\\
Prion & Partial ($N_c=1.0\times10^2$)  &  $3.35\times10^2\pm	1.58\times10^{1}$           &$ \red{1.93\times 10^{-4}}$ & $ \red{7.84\times 10{-2}} $\\
Prion & Partial ($N_c=3.0\times10^1$)  &  $1.69\times10^2\pm	3.34\times10^{0}$           &$ \red{1.81\times 10^{-4}}$ & $ \red{1.08\times 10{-1}} $\\
Prion & Partial ($N_c=1.0\times10^1$)  &  $4.76\times10^1\pm	3.57\times10^{0}$           &$ \red{5.44\times 10^{-4}}$ & $ \red{1.94\times 10{-1}} $\\
\hline
TCR & None (Gillespie SSA) &  $1.04\times10^2\pm	2.27\times10^1$         &$ 0.00\times 10^{0} $ & $ \red{3.06\times 10^{-1}} $\\
TCR & Standard ($\lambda=0.3$)  &  $2.06\times10^1\pm 3.82\times10^0$            &$ \red{2.82\times 10^{-1}}$ & $ \red{2.75\times 10^{-1}} $\\
TCR & Standard ($\lambda=0.1$)  &  $4.86\times10^0\pm	9.44\times10^{-1}$      &$ \red{4.40\times 10^{-1}}$ & $\red{2.29\times 10^{-1}} $\\
TCR & Partial ($N_c=1.0\times10^3$)  &  $3.62\times10^1\pm	5.02\times10^0$                 &$ \red{1.19\times 10^{-3}}$ & $\red{3.16\times 10^{-1}} $\\
TCR & Partial ($N_c=3.0\times10^2$)  &  $1.63\times10^1\pm	1.58\times10^{0}$               &$ \red{1.20\times 10^{-3}}$ & $ \red{2.99\times 10^{-1}} $\\
TCR & Partial ($N_c=1.0\times10^2$)  &  $8.48\times10^0\pm	8.25\times10^{-1}$              &$ \red{9.46\times 10^{-4}}$ & $ \red{3.10\times 10^{-1}} $\\
TCR & Partial ($N_c=3.0\times10^1$)  &  $3.23\times10^0\pm	5.42\times10^{-1}$              &$ \red{1.63\times 10^{-3}}$ & $ \red{3.05\times 10^{-1}} $\\
\end{tabular}
\caption{Efficiency and accuracy of standard scaling and partial scaling methods for stochastic simulation based on models for ERK activation \cite{kochanczyk2017relaxation}, prion protein aggregation \cite{rubenstein2007dynamics}, and TCR signaling \cite{lipniacki2008stochastic}. CPU time was measured using the C++ {\tt clock()} function. \red{Summary statistics for first-moment errors, $\l\langle \varepsilon_1 \r\rangle$, and Van Valen coefficients of variation, $\l\langle CV_\text{VV} \r\rangle$, were calculated as described in the Appendix.}}
\label{table:accuracyEfficiency}
\end{table*}

\section{Discussion and conclusion}

The procedure that we have called standard scaling is commonly used for accelerating stochastic simulations of systems with large population sizes. However, this method has limited applicability and it may be challenging to provide an appropriate/optimal setting for the method's one parameter, a universal static scaling factor $\lambda$. A simulation speed up can only be attained by sacrificing the ability to calculate unbiased estimates of moments higher than the first, and even first-moment information can be calculated incorrectly with overly aggressive scaling. Furthermore, the speed up that is attainable is constrained by the smallest population size measured over the entire time window of simulation. Thus, in practice, careful application of the method may require trial-and-error numerical experiments to find a suitable best setting for the scaling factor $\lambda$. For this reason, the method is especially problematic when one wishes to use it as part of a parameter identification procedure, or any procedure involving parameter variation, because the best $\lambda$ setting depends in a non-obvious way on parameter values (initial conditions and rate constants).

As we have seen for three non-trivial benchmark problems (Figs.~\ref{fig:ERK}--\red{\ref{fig:errors}}, Table~\ref{table:accuracyEfficiency}), partial scaling significantly outperforms standard scaling in two important ways. First, partial scaling yields greater acceleration for the same or better accuracy (\red{Fig.~\ref{fig:errors} and} Table~\ref{table:accuracyEfficiency}). Second, as illustrated in Fig.~\ref{fig:TCRdistribution}, partial scaling better preserves second-moment information (Table~\ref{table:accuracyEfficiency}). Another attractive feature of partial scaling is the better ability of the user to avoid overly aggressive scaling without any requirement for numerical experiments. With partial scaling, overly aggressive scaling is only possible if the setting for $N_c$ is near or below 1, which is not recommended. In contrast, with standard scaling, any setting for $\lambda$ must be tested in numerical experiments. As noted earlier, a trial-and-error procedure may be required to find a suitable setting for $\lambda$ and the suitability of a setting may change with a change of parameter values. Thus, standard scaling seems especially disadvantaged in comparison to partial scaling for simulations within the context of a parameter identification procedure (i.e., a fitting procedure).

\red{Partial scaling is similar to the probability-weighted dynamic Monte Carlo (PW-DMC) method\cite{resat2001probability}. In this method, reaction rates are scaled dynamically according to user-specified rules. The rules define scaling factors for rates that fall into specified ranges. Relative to partial scaling, PW-DMC has more parameters: a set of ranges and the associated scaling factors vs. partial scaling's single parameter, a critical population size ($N_c$). Unlike $N_c$, the PW-DMC parameters must be set through a trial-and-error process. Because rates are scaled without regard to reactant population sizes, PW-DMC can introduce errors where partial scaling does not. These errors arise, for example, when a bimolecular reaction has reactants with disparate population sizes, such that one is discrete but the other is large enough to yield a rate that qualifies for scaling. Scaling in this scenario can cause the discrete population to become negative.}

We have not directly compared partial scaling against hybrid methods or $\tau$-leaping methods. However, partial scaling has clear advantages over these methods. Partial scaling is far easier to implement than hybrid methods, and partial scaling is more broadly applicable than $\tau$-leaping methods. Partial scaling is useful when population sizes are distributed smoothly across multiple scales (i.e., without clear separation of discrete and continuous populations), because there is no \textit{a priori} requirement for classification of population sizes as either discrete or continuous. Such a requirement can be highly problematic when the classification changes during the time window of simulation, as could be the case for a system that exhibits oscillatory behavior. The chief disadvantage of $\tau$-leaping methods is the lack of guarantee of the existence of a time window $\tau$ having the necessary properties. In our experience, hybrid and $\tau$-leaping methods are useful in restricted circumstances. We expect partial scaling to be useful for a wider array of circumstances.

Although partial scaling is an approximate SSA (vs. an exact SSA), such methods have important applications. For example, in parameter identification procedures, which typically entail numerous simulation runs, success is highly dependent on simulation efficiency and inexactness is a lesser concern, especially when parameter estimation is based on noisy data. The approximations of partial scaling are very likely to be tolerable, even when statistical distribution data are being used in a fitting procedure. As the optimization algorithm in a fitting procedure converges, numerous samples around a local optimum are inevitably generated. By averaging over a moving  window of trial parameter sets, for example, one can, in principle, compensate for the noisy (but unbiased) estimates of first moments when determining goodness of fit.

 \red{An attractive future use of partial scaling is acceleration of network-free simulation\red{\cite{Suderman2018}}, which is an important outstanding problem, as we have discussed elsewhere\red{\cite{lin2018using}}. One reason to pursue this goal is the high cost of fitting when network-free simulation is necessary\cite{Thomas2016}, as is the case for the model of Chylek et al.\cite{Chylek2014TCR}.} Unfortunately, application of partial scaling requires system state to be tracked in terms of populations and no currently available, general-purpose implementation of a network-free simulation algorithm tracks system state in this way. Rather, state is followed in terms of the states of individual biomolecular sites\cite{Suderman2018}, which reflects an agent-based or particle-based approach to simulation. However, as discussed by Liu et al. \cite{Liu2010multistate}, it is feasible to develop a network-free simulation algorithm in which system state is tracked in terms of populations, just those of the chemical species with non-zero population sizes. (An exhaustive enumeration of potentially populated chemical species is typically impracticable whenever network-free simulation is under consideration.) Liu et al. \cite{Liu2010multistate} presented a specialized algorithm that uses this approach to state tracking, which they termed the full-scale SSA method. However, this method is specialized for a subset of rule-based models---it doesn't consider systems in which molecules interact to form assemblies (complexes). Thus, an interesting future research direction would be developing a more general version of the full-scale SSA method that is compatible with partial scaling.

\section*{Acknowledgments}
We thank Danny Perez and Arthur F. Voter for helpful discussions. We also thank the research group of James R. Faeder, which maintains the BioNetGen code base with support from the National Institute of General Medical Sciences (NIGMS) of the National Institutes of Health (NIH) (grant no. P41GM103712). W.S.H acknowledges support from NIGMS/NIH (grant no. R01GM111510). Y.T.L. and S.F. acknowledge support from the Center for Nonlinear Studies (CNLS). CNLS is funded by the Laboratory-Directed Research and Development program at Los Alamos National Laboratory, which is operated by Triad National Security, LLC for the National Nuclear Security Administration of the U.S. Department of Energy (contract no. 89233218CNA000001).

\red{

\section*{Appendix}

First-moment errors of scaled simulations per species and overall are reported in Fig.~\ref{fig:errors} and Table~\ref{table:accuracyEfficiency}, respectively. These errors were calculated as described here. Below, we also describe the means used to compare the aggressiveness of standard and partial scaling fairly (via calculation of an effective scaling factor, $\lambda_{\rm eff}$) and the means used to characterize second moments obtained from either an unscaled or scaled simulation in terms of a multivariate coefficient of variation.

For simulations based on any of the three models of interest involving scaling (either standard scaling or partial scaling), the mean population of chemical species $X_i$ (for $i=1,\ldots,M$) at report time $t$ was calculated from $n_s$ sample paths. Similarly, on the basis of $n_s$ sample paths, we calculated the corresponding mean obtained via exact simulation. In this way, for either scaling scheme of interest (standard scaling with $0<\lambda<1$ or partial scaling with $N_c \gg 1$), we obtain a deviation vector
\eq{
\Delta \overline{\mathbf{N}^{(\lambda,N_c)}}\l(t\r) \equiv \overline{\mathbf{N}^{(\lambda,N_c)}}\l(t\r) - \overline{\mathbf{N}^{(0)}}\l(t\r).
}{}
where $\overline{\mathbf{N}^{(\lambda,N_c)}}\l(t\r)$ denotes the time-dependent $M$-dimensional vector of sample path means calculated from scaled simulations and $\overline{\mathbf{N}^{(0)}}(t)$ denotes the time-dependent $M$-dimensional vector of sample path means calculated from exact simulations. In using the notation $\overline{\mathbf{N}^{(\lambda,N_c)}}$, we adopt the convention that replacement of $(\lambda, N_c)$ with $(1, \infty)$, $(\lambda, \infty)$ and $(1, N_c)$ indicates no scaling, standard scaling, and partial scaling, respectively.

To quantify the accuracy of first moments obtained from scaled simulations, we define the following error measure:
\eq{
\varepsilon^i \l(t\r) \equiv \l\{ 
\begin{array}{l}
\frac{\bar{N}_i^{\l(\lambda, \bar{N}_c\r)} \l(t\r)-\bar{N}_i^{\l(0\r)}\l(t\r)}{\bar{N}_i^{\l(0\r)}\l(t\r)}\text{, if } \bar{N}_i\l(t\r)\ge1,\\
\bar{N}_i^{\l(\lambda, \bar{N}_c\r)}\l(t\r)-\bar{N}_i^{\l(0\r)}\l(t\r)\text{, otherwise}
\end{array}
\r.
}{} \label{eq:error_metric}
Here, $\bar{N}_i\l(t\r)$ denotes the averaged population of species $i$ at time $t$ determined from a set of unscaled simulations. Similarly, $\bar{N}_i^{\l(\lambda, \bar{N}_c\r)}\l(t\r)$ denotes the corresponding quantity determined from an equal number of scaled simulations.
We note that $\varepsilon_1^i$ is continuous with respect to $\bar{N}_i^{\l(\lambda, N_c\r)}$ for arbitrary $\bar{N}_i \geq 0$. The derivative of $\varepsilon_1^i$ with respect to  $\bar{N}_i^{\l(\lambda, N_c\r)}$ is also continuous. With the definition given above, error does not diverge to infinity as $\bar{N}_i$ approaches 0.

For each numerical experiment of interest (aimed at quantifying the errors introduced by scaling), we calculated $\varepsilon_1^i$ for each $i$ at a set of evenly spaced report times, from 0 to $T$. We then calculated the temporal average $\l\langle \varepsilon_1^i \r\rangle$, which is given by
\eq{
\l\langle \varepsilon^i \r\rangle = \frac{\Delta t}{T} \sum_{k=0}^{T/\Delta t} \varepsilon^{i} \l(k\Delta t\r)
}{}
where $\Delta t$ is the spacing between report times. Quantities found using this equation are plotted in  Fig.~\ref{fig:errors}.

The aggressiveness of partial scaling is related to $N_c$, and the aggressiveness of standard scaling is related to $\lambda$. These quantities are not directly comparable. To fairly compare the aggressiveness of partial and standard scaling, we define $\lambda_\text{eff}$ as the normalized value of $\kappa$ in a scaled simulation (Algorithm \ref{alg:partialScaling}, line 14) averaged over all discrete report times. We take the normalization constant to be the temporally averaged value of $\kappa$ in a corresponding unscaled simulation. Thus, in the case of standard scaling, $\lambda_\text{eff}$ is exactly $\lambda$. In Fig.~\ref{fig:errors}, we indicate the values of $\lambda_\text{eff}$ calculated for partially scaled simulations with different values of $N_c$.

In Table \ref{table:accuracyEfficiency}, for each numerical experiment, we report a summary statistic for overall error, which is defined as
\eq{
\l\langle \varepsilon \r\rangle \equiv \sum_{i=1}^M \l\langle \varepsilon^i \r\rangle 
}{}

To characterize second moments, we calculated the Van Valen multivariate coefficient of variation\cite{van1974multivariate} of the $M$-dimensional vector $\mathbf{N}(t)$ describing the species popualtions at each report time:
\eq{
CV_\text{VV}\l(t\r) \equiv \sqrt{\frac{ \text{tr}\l(\Sigma \l(t\r) \r) }{\l \vert\vert \bar{\mathbf{N}}\l(t\r) \vert \r \vert_2^2}},
}{}
where $\l \vert\vert \bar{\mathbf{N}}\l(t\r) \vert \r \vert_2^2$ is the square of the $L^2$ norm of the averaged populations and $\Sigma \l(t\r)$ is the covariance matrix. To report a second-moment summary statistic, we averaged over report times:
\eq{
\l\langle CV_\text{VV} \r\rangle =  \frac{\Delta t}{T} \sum_{k=0}^{T/\Delta t} CV_\text{VV}\l(k \Delta t\r).
}{}
Values of $\l\langle CV_\text{VV} \r\rangle$ are reported in Table~\ref{table:accuracyEfficiency} for both scaled and unscaled simulations. Second-moment errors attributable to scaling are reflected in the difference between the values of $\l\langle CV_\text{VV} \r\rangle$ given for a scaled simulation and the corresponding unscaled simulation.

Values given in Table~\ref{table:accuracyEfficiency} for $\l\langle \varepsilon_1 \r\rangle$ and $\l\langle CV_\text{VV} \r\rangle$ are based on the following choices for $\Delta t$, $T$, and $n_s$, the number of sample paths. For the ERK activation model\cite{kochanczyk2017relaxation}, $\Delta t = 8.64$ sec, $T=8,640$ sec, and $n_s=500$. For the prion protein aggregation model\cite{rubenstein2007dynamics}, $\Delta t = 0.01$ day, $T=300$ days, and $n_s=500$. For the TCR signaling model\cite{lipniacki2008stochastic}, $\Delta t =10$ sec, $T=10^4$ sec, and $n_s=10^4$.

} 

\let\l\polishl


\begin{thebibliography}{65}%
\makeatletter
\providecommand \@ifxundefined [1]{%
 \@ifx{#1\undefined}
}%
\providecommand \@ifnum [1]{%
 \ifnum #1\expandafter \@firstoftwo
 \else \expandafter \@secondoftwo
 \fi
}%
\providecommand \@ifx [1]{%
 \ifx #1\expandafter \@firstoftwo
 \else \expandafter \@secondoftwo
 \fi
}%
\providecommand \natexlab [1]{#1}%
\providecommand \enquote  [1]{``#1''}%
\providecommand \bibnamefont  [1]{#1}%
\providecommand \bibfnamefont [1]{#1}%
\providecommand \citenamefont [1]{#1}%
\providecommand \href@noop [0]{\@secondoftwo}%
\providecommand \href [0]{\begingroup \@sanitize@url \@href}%
\providecommand \@href[1]{\@@startlink{#1}\@@href}%
\providecommand \@@href[1]{\endgroup#1\@@endlink}%
\providecommand \@sanitize@url [0]{\catcode `\\12\catcode `\$12\catcode
  `\&12\catcode `\#12\catcode `\^12\catcode `\_12\catcode `\%12\relax}%
\providecommand \@@startlink[1]{}%
\providecommand \@@endlink[0]{}%
\providecommand \url  [0]{\begingroup\@sanitize@url \@url }%
\providecommand \@url [1]{\endgroup\@href {#1}{\urlprefix }}%
\providecommand \urlprefix  [0]{URL }%
\providecommand \Eprint [0]{\href }%
\providecommand \doibase [0]{http://dx.doi.org/}%
\providecommand \selectlanguage [0]{\@gobble}%
\providecommand \bibinfo  [0]{\@secondoftwo}%
\providecommand \bibfield  [0]{\@secondoftwo}%
\providecommand \translation [1]{[#1]}%
\providecommand \BibitemOpen [0]{}%
\providecommand \bibitemStop [0]{}%
\providecommand \bibitemNoStop [0]{.\EOS\space}%
\providecommand \EOS [0]{\spacefactor3000\relax}%
\providecommand \BibitemShut  [1]{\csname bibitem#1\endcsname}%
\let\auto@bib@innerbib\@empty
\bibitem [{\citenamefont {Voter}(2007)}]{voter2007introduction}%
  \BibitemOpen
  \bibfield  {author} {\bibinfo {author} {\bibfnamefont {A.~F.}\ \bibnamefont
  {Voter}},\ }\bibfield  {title} {\enquote {\bibinfo {title} {Introduction to
  the kinetic {Monte Carlo} method},}\ }in\ \href@noop {} {\emph {\bibinfo
  {booktitle} {Radiation effects in solids}}}\ (\bibinfo  {publisher}
  {Springer},\ \bibinfo {year} {2007})\ pp.\ \bibinfo {pages}
  {1--23}\BibitemShut {NoStop}%
\bibitem [{\citenamefont {Gillespie}(2007)}]{Gillespie2007}%
  \BibitemOpen
  \bibfield  {author} {\bibinfo {author} {\bibfnamefont {D.~T.}\ \bibnamefont
  {Gillespie}},\ }\bibfield  {title} {\enquote {\bibinfo {title} {Stochastic
  simulation of chemical kinetics},}\ }\href@noop {} {\bibfield  {journal}
  {\bibinfo  {journal} {Annu Rev Phys Chem}\ }\textbf {\bibinfo {volume}
  {58}},\ \bibinfo {pages} {35--55} (\bibinfo {year} {2007})}\BibitemShut
  {NoStop}%
\bibitem [{\citenamefont {McAdams}\ and\ \citenamefont
  {Arkin}(1999)}]{McAdams1999}%
  \BibitemOpen
  \bibfield  {author} {\bibinfo {author} {\bibfnamefont {H.~H.}\ \bibnamefont
  {McAdams}}\ and\ \bibinfo {author} {\bibfnamefont {A.}~\bibnamefont
  {Arkin}},\ }\bibfield  {title} {\enquote {\bibinfo {title} {It's a noisy
  business! {Genetic} regulation at the nanomolar scale},}\ }\href@noop {}
  {\bibfield  {journal} {\bibinfo  {journal} {Trends Genet}\ }\textbf {\bibinfo
  {volume} {15}},\ \bibinfo {pages} {65--69} (\bibinfo {year}
  {1999})}\BibitemShut {NoStop}%
\bibitem [{\citenamefont {Suderman}\ \emph {et~al.}(2018)\citenamefont
  {Suderman}, \citenamefont {Mitra}, \citenamefont {Lin}, \citenamefont
  {Erickson}, \citenamefont {Feng},\ and\ \citenamefont
  {Hlavacek}}]{Suderman2018}%
  \BibitemOpen
  \bibfield  {author} {\bibinfo {author} {\bibfnamefont {R.}~\bibnamefont
  {Suderman}}, \bibinfo {author} {\bibfnamefont {E.~D.}\ \bibnamefont {Mitra}},
  \bibinfo {author} {\bibfnamefont {Y.~T.}\ \bibnamefont {Lin}}, \bibinfo
  {author} {\bibfnamefont {K.~E.}\ \bibnamefont {Erickson}}, \bibinfo {author}
  {\bibfnamefont {S.}~\bibnamefont {Feng}}, \ and\ \bibinfo {author}
  {\bibfnamefont {W.~S.}\ \bibnamefont {Hlavacek}},\ }\bibfield  {title}
  {\enquote {\bibinfo {title} {Generalizing {Gillespie}'s direct method to
  enable network-free simulations},}\ }\href
  {https://doi.org/10.1007/s11538-018-0418-2} {\bibfield  {journal} {\bibinfo
  {journal} {Bull Math Biol}\ } (\bibinfo {year} {2018})}\BibitemShut {NoStop}%
\bibitem [{\citenamefont {Chylek}\ \emph
  {et~al.}(2014{\natexlab{a}})\citenamefont {Chylek}, \citenamefont {Harris},
  \citenamefont {Tung}, \citenamefont {Faeder}, \citenamefont {Lopez},\ and\
  \citenamefont {Hlavacek}}]{Chylek2014}%
  \BibitemOpen
  \bibfield  {author} {\bibinfo {author} {\bibfnamefont {L.~A.}\ \bibnamefont
  {Chylek}}, \bibinfo {author} {\bibfnamefont {L.~A.}\ \bibnamefont {Harris}},
  \bibinfo {author} {\bibfnamefont {C.-S.}\ \bibnamefont {Tung}}, \bibinfo
  {author} {\bibfnamefont {J.~R.}\ \bibnamefont {Faeder}}, \bibinfo {author}
  {\bibfnamefont {C.~F.}\ \bibnamefont {Lopez}}, \ and\ \bibinfo {author}
  {\bibfnamefont {W.~S.}\ \bibnamefont {Hlavacek}},\ }\bibfield  {title}
  {\enquote {\bibinfo {title} {Rule-based modeling: a computational approach
  for studying biomolecular site dynamics in cell signaling systems},}\
  }\href@noop {} {\bibfield  {journal} {\bibinfo  {journal} {Wiley Interdiscip
  Rev Syst Biol Med}\ }\textbf {\bibinfo {volume} {6}},\ \bibinfo {pages}
  {13--36} (\bibinfo {year} {2014}{\natexlab{a}})}\BibitemShut {NoStop}%
\bibitem [{\citenamefont {Yang}\ \emph {et~al.}(2008)\citenamefont {Yang},
  \citenamefont {Monine}, \citenamefont {Faeder},\ and\ \citenamefont
  {Hlavacek}}]{Yang2008}%
  \BibitemOpen
  \bibfield  {author} {\bibinfo {author} {\bibfnamefont {J.}~\bibnamefont
  {Yang}}, \bibinfo {author} {\bibfnamefont {M.~I.}\ \bibnamefont {Monine}},
  \bibinfo {author} {\bibfnamefont {J.~R.}\ \bibnamefont {Faeder}}, \ and\
  \bibinfo {author} {\bibfnamefont {W.~S.}\ \bibnamefont {Hlavacek}},\
  }\bibfield  {title} {\enquote {\bibinfo {title} {Kinetic {Monte Carlo} method
  for rule-based modeling of biochemical networks},}\ }\href@noop {} {\bibfield
   {journal} {\bibinfo  {journal} {Phys Rev E}\ }\textbf {\bibinfo {volume}
  {78}},\ \bibinfo {pages} {031910} (\bibinfo {year} {2008})}\BibitemShut
  {NoStop}%
\bibitem [{\citenamefont {Sneddon}, \citenamefont {Faeder},\ and\ \citenamefont
  {Emonet}(2011)}]{Sneddon2011}%
  \BibitemOpen
  \bibfield  {author} {\bibinfo {author} {\bibfnamefont {M.~W.}\ \bibnamefont
  {Sneddon}}, \bibinfo {author} {\bibfnamefont {J.~R.}\ \bibnamefont {Faeder}},
  \ and\ \bibinfo {author} {\bibfnamefont {T.}~\bibnamefont {Emonet}},\
  }\bibfield  {title} {\enquote {\bibinfo {title} {Efficient modeling,
  simulation and coarse-graining of biological complexity with {NFsim}},}\
  }\href@noop {} {\bibfield  {journal} {\bibinfo  {journal} {Nat Methods}\
  }\textbf {\bibinfo {volume} {8}},\ \bibinfo {pages} {177--183} (\bibinfo
  {year} {2011})}\BibitemShut {NoStop}%
\bibitem [{\citenamefont {Creamer}\ \emph {et~al.}(2012)\citenamefont
  {Creamer}, \citenamefont {Stites}, \citenamefont {Aziz}, \citenamefont
  {Cahill}, \citenamefont {Tan}, \citenamefont {Berens}, \citenamefont {Han},
  \citenamefont {Bussey}, \citenamefont {Von~Hoff}, \citenamefont {Hlavacek},\
  and\ \citenamefont {Posner}}]{Creamer2012}%
  \BibitemOpen
  \bibfield  {author} {\bibinfo {author} {\bibfnamefont {M.~S.}\ \bibnamefont
  {Creamer}}, \bibinfo {author} {\bibfnamefont {E.~C.}\ \bibnamefont {Stites}},
  \bibinfo {author} {\bibfnamefont {M.}~\bibnamefont {Aziz}}, \bibinfo {author}
  {\bibfnamefont {J.~A.}\ \bibnamefont {Cahill}}, \bibinfo {author}
  {\bibfnamefont {C.~W.}\ \bibnamefont {Tan}}, \bibinfo {author} {\bibfnamefont
  {M.~E.}\ \bibnamefont {Berens}}, \bibinfo {author} {\bibfnamefont
  {H.}~\bibnamefont {Han}}, \bibinfo {author} {\bibfnamefont {K.~J.}\
  \bibnamefont {Bussey}}, \bibinfo {author} {\bibfnamefont {D.~D.}\
  \bibnamefont {Von~Hoff}}, \bibinfo {author} {\bibfnamefont {W.~S.}\
  \bibnamefont {Hlavacek}}, \ and\ \bibinfo {author} {\bibfnamefont {R.~G.}\
  \bibnamefont {Posner}},\ }\bibfield  {title} {\enquote {\bibinfo {title}
  {Specification, annotation, visualization and simulation of a large
  rule-based model for {ERBB} receptor signaling},}\ }\href@noop {} {\bibfield
  {journal} {\bibinfo  {journal} {BMC Syst Biol}\ }\textbf {\bibinfo {volume}
  {6}},\ \bibinfo {pages} {107} (\bibinfo {year} {2012})}\BibitemShut {NoStop}%
\bibitem [{\citenamefont {Van~Kampen}(1992)}]{van1992stochastic}%
  \BibitemOpen
  \bibfield  {author} {\bibinfo {author} {\bibfnamefont {N.~G.}\ \bibnamefont
  {Van~Kampen}},\ }\href@noop {} {\emph {\bibinfo {title} {Stochastic processes
  in physics and chemistry}}},\ Vol.~\bibinfo {volume} {1}\ (\bibinfo
  {publisher} {Elsevier},\ \bibinfo {year} {1992})\BibitemShut {NoStop}%
\bibitem [{\citenamefont {Risken}(1996)}]{risken1996fokker}%
  \BibitemOpen
  \bibfield  {author} {\bibinfo {author} {\bibfnamefont {H.}~\bibnamefont
  {Risken}},\ }\bibfield  {title} {\enquote {\bibinfo {title} {Fokker-{P}lanck
  equation},}\ }in\ \href@noop {} {\emph {\bibinfo {booktitle} {The
  {Fokker-Planck} Equation}}}\ (\bibinfo  {publisher} {Springer},\ \bibinfo
  {year} {1996})\ pp.\ \bibinfo {pages} {63--95}\BibitemShut {NoStop}%
\bibitem [{\citenamefont {Gillespie}(2001)}]{gillespie2001approximate}%
  \BibitemOpen
  \bibfield  {author} {\bibinfo {author} {\bibfnamefont {D.~T.}\ \bibnamefont
  {Gillespie}},\ }\bibfield  {title} {\enquote {\bibinfo {title} {Approximate
  accelerated stochastic simulation of chemically reacting systems},}\
  }\href@noop {} {\bibfield  {journal} {\bibinfo  {journal} {The Journal of
  Chemical Physics}\ }\textbf {\bibinfo {volume} {115}},\ \bibinfo {pages}
  {1716--1733} (\bibinfo {year} {2001})}\BibitemShut {NoStop}%
\bibitem [{\citenamefont {Salis}\ and\ \citenamefont
  {Kaznessis}(2005)}]{Salis05accurate}%
  \BibitemOpen
  \bibfield  {author} {\bibinfo {author} {\bibfnamefont {H.}~\bibnamefont
  {Salis}}\ and\ \bibinfo {author} {\bibfnamefont {Y.}~\bibnamefont
  {Kaznessis}},\ }\bibfield  {title} {\enquote {\bibinfo {title} {Accurate
  hybrid stochastic simulation of a system of coupled chemical or biochemical
  reactions},}\ }\href {\doibase 10.1063/1.1835951} {\bibfield  {journal}
  {\bibinfo  {journal} {The Journal of Chemical Physics}\ }\textbf {\bibinfo
  {volume} {122}},\ \bibinfo {pages} {054103} (\bibinfo {year}
  {2005})}\BibitemShut {NoStop}%
\bibitem [{\citenamefont {Newby}, \citenamefont {Bressloff},\ and\
  \citenamefont {Keener}(2013)}]{newby2013breakdown}%
  \BibitemOpen
  \bibfield  {author} {\bibinfo {author} {\bibfnamefont {J.~M.}\ \bibnamefont
  {Newby}}, \bibinfo {author} {\bibfnamefont {P.~C.}\ \bibnamefont
  {Bressloff}}, \ and\ \bibinfo {author} {\bibfnamefont {J.~P.}\ \bibnamefont
  {Keener}},\ }\bibfield  {title} {\enquote {\bibinfo {title} {Breakdown of
  fast-slow analysis in an excitable system with channel noise},}\ }\href@noop
  {} {\bibfield  {journal} {\bibinfo  {journal} {Physical review letters}\
  }\textbf {\bibinfo {volume} {111}},\ \bibinfo {pages} {128101} (\bibinfo
  {year} {2013})}\BibitemShut {NoStop}%
\bibitem [{\citenamefont {Bokes}\ \emph {et~al.}(2013)\citenamefont {Bokes},
  \citenamefont {King}, \citenamefont {Wood},\ and\ \citenamefont
  {Loose}}]{bokes2013transcriptional}%
  \BibitemOpen
  \bibfield  {author} {\bibinfo {author} {\bibfnamefont {P.}~\bibnamefont
  {Bokes}}, \bibinfo {author} {\bibfnamefont {J.~R.}\ \bibnamefont {King}},
  \bibinfo {author} {\bibfnamefont {A.~T.}\ \bibnamefont {Wood}}, \ and\
  \bibinfo {author} {\bibfnamefont {M.}~\bibnamefont {Loose}},\ }\bibfield
  {title} {\enquote {\bibinfo {title} {Transcriptional bursting diversifies the
  behaviour of a toggle switch: hybrid simulation of stochastic gene
  expression},}\ }\href@noop {} {\bibfield  {journal} {\bibinfo  {journal}
  {Bulletin of mathematical biology}\ }\textbf {\bibinfo {volume} {75}},\
  \bibinfo {pages} {351--371} (\bibinfo {year} {2013})}\BibitemShut {NoStop}%
\bibitem [{\citenamefont {Bressloff}\ and\ \citenamefont
  {Newby}(2014)}]{bressloff2014stochastic}%
  \BibitemOpen
  \bibfield  {author} {\bibinfo {author} {\bibfnamefont {P.~C.}\ \bibnamefont
  {Bressloff}}\ and\ \bibinfo {author} {\bibfnamefont {J.~M.}\ \bibnamefont
  {Newby}},\ }\bibfield  {title} {\enquote {\bibinfo {title} {Stochastic hybrid
  model of spontaneous dendritic {NMDA} spikes},}\ }\href@noop {} {\bibfield
  {journal} {\bibinfo  {journal} {Physical biology}\ }\textbf {\bibinfo
  {volume} {11}},\ \bibinfo {pages} {016006} (\bibinfo {year}
  {2014})}\BibitemShut {NoStop}%
\bibitem [{\citenamefont {Bressloff}(2015)}]{bressloff2015path}%
  \BibitemOpen
  \bibfield  {author} {\bibinfo {author} {\bibfnamefont {P.~C.}\ \bibnamefont
  {Bressloff}},\ }\bibfield  {title} {\enquote {\bibinfo {title} {Path-integral
  methods for analyzing the effects of fluctuations in stochastic hybrid neural
  networks},}\ }\href@noop {} {\bibfield  {journal} {\bibinfo  {journal} {The
  Journal of Mathematical Neuroscience (JMN)}\ }\textbf {\bibinfo {volume}
  {5}},\ \bibinfo {pages} {4} (\bibinfo {year} {2015})}\BibitemShut {NoStop}%
\bibitem [{\citenamefont {Lin}\ and\ \citenamefont
  {Doering}(2016)}]{lin2016gene}%
  \BibitemOpen
  \bibfield  {author} {\bibinfo {author} {\bibfnamefont {Y.~T.}\ \bibnamefont
  {Lin}}\ and\ \bibinfo {author} {\bibfnamefont {C.~R.}\ \bibnamefont
  {Doering}},\ }\bibfield  {title} {\enquote {\bibinfo {title} {Gene expression
  dynamics with stochastic bursts: Construction and exact results for a
  coarse-grained model},}\ }\href@noop {} {\bibfield  {journal} {\bibinfo
  {journal} {Physical Review E}\ }\textbf {\bibinfo {volume} {93}},\ \bibinfo
  {pages} {022409} (\bibinfo {year} {2016})}\BibitemShut {NoStop}%
\bibitem [{\citenamefont {Lin}\ and\ \citenamefont
  {Galla}(2016)}]{lin2016bursting}%
  \BibitemOpen
  \bibfield  {author} {\bibinfo {author} {\bibfnamefont {Y.~T.}\ \bibnamefont
  {Lin}}\ and\ \bibinfo {author} {\bibfnamefont {T.}~\bibnamefont {Galla}},\
  }\bibfield  {title} {\enquote {\bibinfo {title} {Bursting noise in gene
  expression dynamics: linking microscopic and mesoscopic models},}\
  }\href@noop {} {\bibfield  {journal} {\bibinfo  {journal} {Journal of The
  Royal Society Interface}\ }\textbf {\bibinfo {volume} {13}},\ \bibinfo
  {pages} {20150772} (\bibinfo {year} {2016})}\BibitemShut {NoStop}%
\bibitem [{\citenamefont
  {Bressloff}(2017{\natexlab{a}})}]{bressloff2017stochastic}%
  \BibitemOpen
  \bibfield  {author} {\bibinfo {author} {\bibfnamefont {P.~C.}\ \bibnamefont
  {Bressloff}},\ }\bibfield  {title} {\enquote {\bibinfo {title} {Stochastic
  switching in biology: from genotype to phenotype},}\ }\href@noop {}
  {\bibfield  {journal} {\bibinfo  {journal} {Journal of Physics A:
  Mathematical and Theoretical}\ }\textbf {\bibinfo {volume} {50}},\ \bibinfo
  {pages} {133001} (\bibinfo {year} {2017}{\natexlab{a}})}\BibitemShut
  {NoStop}%
\bibitem [{\citenamefont
  {Bressloff}(2017{\natexlab{b}})}]{bressloff2017feynman}%
  \BibitemOpen
  \bibfield  {author} {\bibinfo {author} {\bibfnamefont {P.~C.}\ \bibnamefont
  {Bressloff}},\ }\bibfield  {title} {\enquote {\bibinfo {title} {Feynman-{Kac}
  formula for stochastic hybrid systems},}\ }\href@noop {} {\bibfield
  {journal} {\bibinfo  {journal} {Physical Review E}\ }\textbf {\bibinfo
  {volume} {95}},\ \bibinfo {pages} {012138} (\bibinfo {year}
  {2017}{\natexlab{b}})}\BibitemShut {NoStop}%
\bibitem [{\citenamefont {Lin}\ and\ \citenamefont
  {Buchler}(2018)}]{lin2018efficient}%
  \BibitemOpen
  \bibfield  {author} {\bibinfo {author} {\bibfnamefont {Y.~T.}\ \bibnamefont
  {Lin}}\ and\ \bibinfo {author} {\bibfnamefont {N.~E.}\ \bibnamefont
  {Buchler}},\ }\bibfield  {title} {\enquote {\bibinfo {title} {Efficient
  analysis of stochastic gene dynamics in the non-adiabatic regime using
  piecewise deterministic {Markov} processes},}\ }\href@noop {} {\bibfield
  {journal} {\bibinfo  {journal} {Journal of The Royal Society Interface}\
  }\textbf {\bibinfo {volume} {15}},\ \bibinfo {pages} {20170804} (\bibinfo
  {year} {2018})}\BibitemShut {NoStop}%
\bibitem [{\citenamefont {Mao}\ and\ \citenamefont
  {Yuan}(2006)}]{mao2006stochastic}%
  \BibitemOpen
  \bibfield  {author} {\bibinfo {author} {\bibfnamefont {X.}~\bibnamefont
  {Mao}}\ and\ \bibinfo {author} {\bibfnamefont {C.}~\bibnamefont {Yuan}},\
  }\href@noop {} {\emph {\bibinfo {title} {Stochastic differential equations
  with Markovian switching}}}\ (\bibinfo  {publisher} {World Scientific},\
  \bibinfo {year} {2006})\BibitemShut {NoStop}%
\bibitem [{\citenamefont {Kepler}\ and\ \citenamefont
  {Elston}(2001)}]{Kepler2001}%
  \BibitemOpen
  \bibfield  {author} {\bibinfo {author} {\bibfnamefont {T.~B.}\ \bibnamefont
  {Kepler}}\ and\ \bibinfo {author} {\bibfnamefont {T.~C.}\ \bibnamefont
  {Elston}},\ }\bibfield  {title} {\enquote {\bibinfo {title} {Stochasticity in
  transcriptional regulation: origins, consequences, and mathematical
  representations},}\ }\href@noop {} {\bibfield  {journal} {\bibinfo  {journal}
  {Biophys J}\ }\textbf {\bibinfo {volume} {81}},\ \bibinfo {pages}
  {3116--3136} (\bibinfo {year} {2001})}\BibitemShut {NoStop}%
\bibitem [{\citenamefont {Faeder}, \citenamefont {Blinov},\ and\ \citenamefont
  {Hlavacek}(2009)}]{Faeder2009}%
  \BibitemOpen
  \bibfield  {author} {\bibinfo {author} {\bibfnamefont {J.~R.}\ \bibnamefont
  {Faeder}}, \bibinfo {author} {\bibfnamefont {M.~L.}\ \bibnamefont {Blinov}},
  \ and\ \bibinfo {author} {\bibfnamefont {W.~S.}\ \bibnamefont {Hlavacek}},\
  }\bibfield  {title} {\enquote {\bibinfo {title} {Rule-based modeling of
  biochemical systems with {BioNetGen}},}\ }\href@noop {} {\bibfield  {journal}
  {\bibinfo  {journal} {Methods Mol Biol}\ }\textbf {\bibinfo {volume} {500}},\
  \bibinfo {pages} {113--167} (\bibinfo {year} {2009})}\BibitemShut {NoStop}%
\bibitem [{\citenamefont {Kocha{\'n}czyk}\ \emph {et~al.}(2017)\citenamefont
  {Kocha{\'n}czyk}, \citenamefont {Kocieniewski}, \citenamefont {Koz{\l}owska},
  \citenamefont {Jaruszewicz-B{\l}o{\'n}ska}, \citenamefont {Sparta},
  \citenamefont {Pargett}, \citenamefont {Albeck}, \citenamefont {Hlavacek},\
  and\ \citenamefont {Lipniacki}}]{kochanczyk2017relaxation}%
  \BibitemOpen
  \bibfield  {author} {\bibinfo {author} {\bibfnamefont {M.}~\bibnamefont
  {Kocha{\'n}czyk}}, \bibinfo {author} {\bibfnamefont {P.}~\bibnamefont
  {Kocieniewski}}, \bibinfo {author} {\bibfnamefont {E.}~\bibnamefont
  {Koz{\l}owska}}, \bibinfo {author} {\bibfnamefont {J.}~\bibnamefont
  {Jaruszewicz-B{\l}o{\'n}ska}}, \bibinfo {author} {\bibfnamefont
  {B.}~\bibnamefont {Sparta}}, \bibinfo {author} {\bibfnamefont
  {M.}~\bibnamefont {Pargett}}, \bibinfo {author} {\bibfnamefont {J.~G.}\
  \bibnamefont {Albeck}}, \bibinfo {author} {\bibfnamefont {W.~S.}\
  \bibnamefont {Hlavacek}}, \ and\ \bibinfo {author} {\bibfnamefont
  {T.}~\bibnamefont {Lipniacki}},\ }\bibfield  {title} {\enquote {\bibinfo
  {title} {Relaxation oscillations and hierarchy of feedbacks in {MAPK}
  signaling},}\ }\href@noop {} {\bibfield  {journal} {\bibinfo  {journal}
  {Scientific reports}\ }\textbf {\bibinfo {volume} {7}},\ \bibinfo {pages}
  {38244} (\bibinfo {year} {2017})}\BibitemShut {NoStop}%
\bibitem [{\citenamefont {Rubenstein}\ \emph {et~al.}(2007)\citenamefont
  {Rubenstein}, \citenamefont {Gray}, \citenamefont {Cleland}, \citenamefont
  {Piltch}, \citenamefont {Hlavacek}, \citenamefont {Roberts}, \citenamefont
  {Ambrosiano},\ and\ \citenamefont {Kim}}]{rubenstein2007dynamics}%
  \BibitemOpen
  \bibfield  {author} {\bibinfo {author} {\bibfnamefont {R.}~\bibnamefont
  {Rubenstein}}, \bibinfo {author} {\bibfnamefont {P.~C.}\ \bibnamefont
  {Gray}}, \bibinfo {author} {\bibfnamefont {T.~J.}\ \bibnamefont {Cleland}},
  \bibinfo {author} {\bibfnamefont {M.~S.}\ \bibnamefont {Piltch}}, \bibinfo
  {author} {\bibfnamefont {W.~S.}\ \bibnamefont {Hlavacek}}, \bibinfo {author}
  {\bibfnamefont {R.~M.}\ \bibnamefont {Roberts}}, \bibinfo {author}
  {\bibfnamefont {J.}~\bibnamefont {Ambrosiano}}, \ and\ \bibinfo {author}
  {\bibfnamefont {J.-I.}\ \bibnamefont {Kim}},\ }\bibfield  {title} {\enquote
  {\bibinfo {title} {Dynamics of the nucleated polymerization model of prion
  replication},}\ }\href@noop {} {\bibfield  {journal} {\bibinfo  {journal}
  {Biophysical chemistry}\ }\textbf {\bibinfo {volume} {125}},\ \bibinfo
  {pages} {360--367} (\bibinfo {year} {2007})}\BibitemShut {NoStop}%
\bibitem [{\citenamefont {Lipniacki}\ \emph {et~al.}(2008)\citenamefont
  {Lipniacki}, \citenamefont {Hat}, \citenamefont {Faeder},\ and\ \citenamefont
  {Hlavacek}}]{lipniacki2008stochastic}%
  \BibitemOpen
  \bibfield  {author} {\bibinfo {author} {\bibfnamefont {T.}~\bibnamefont
  {Lipniacki}}, \bibinfo {author} {\bibfnamefont {B.}~\bibnamefont {Hat}},
  \bibinfo {author} {\bibfnamefont {J.~R.}\ \bibnamefont {Faeder}}, \ and\
  \bibinfo {author} {\bibfnamefont {W.~S.}\ \bibnamefont {Hlavacek}},\
  }\bibfield  {title} {\enquote {\bibinfo {title} {Stochastic effects and
  bistability in {T} cell receptor signaling},}\ }\href@noop {} {\bibfield
  {journal} {\bibinfo  {journal} {Journal of theoretical Biology}\ }\textbf
  {\bibinfo {volume} {254}},\ \bibinfo {pages} {110--122} (\bibinfo {year}
  {2008})}\BibitemShut {NoStop}%
\bibitem [{\citenamefont {Harris}\ \emph {et~al.}(2016)\citenamefont {Harris},
  \citenamefont {Hogg}, \citenamefont {Tapia}, \citenamefont {Sekar},
  \citenamefont {Gupta}, \citenamefont {Korsunsky}, \citenamefont {Arora},
  \citenamefont {Barua}, \citenamefont {Sheehan},\ and\ \citenamefont
  {Faeder}}]{Harris2016}%
  \BibitemOpen
  \bibfield  {author} {\bibinfo {author} {\bibfnamefont {L.~A.}\ \bibnamefont
  {Harris}}, \bibinfo {author} {\bibfnamefont {J.~S.}\ \bibnamefont {Hogg}},
  \bibinfo {author} {\bibfnamefont {J.-J.}\ \bibnamefont {Tapia}}, \bibinfo
  {author} {\bibfnamefont {J.~A.~P.}\ \bibnamefont {Sekar}}, \bibinfo {author}
  {\bibfnamefont {S.}~\bibnamefont {Gupta}}, \bibinfo {author} {\bibfnamefont
  {I.}~\bibnamefont {Korsunsky}}, \bibinfo {author} {\bibfnamefont
  {A.}~\bibnamefont {Arora}}, \bibinfo {author} {\bibfnamefont
  {D.}~\bibnamefont {Barua}}, \bibinfo {author} {\bibfnamefont {R.~P.}\
  \bibnamefont {Sheehan}}, \ and\ \bibinfo {author} {\bibfnamefont {J.~R.}\
  \bibnamefont {Faeder}},\ }\bibfield  {title} {\enquote {\bibinfo {title}
  {{BioNetGen} 2.2: advances in rule-based modeling},}\ }\href@noop {}
  {\bibfield  {journal} {\bibinfo  {journal} {Bioinformatics}\ }\textbf
  {\bibinfo {volume} {32}},\ \bibinfo {pages} {3366--3368} (\bibinfo {year}
  {2016})}\BibitemShut {NoStop}%
\bibitem [{Note1()}]{Note1}%
  \BibitemOpen
  \bibinfo {note} {The parameter $\Omega $ characterizes system size. Here, we
  take it to represent volume but it can alternatively be interpreted as a
  population that defines a population scale. With this interpretation,
  (dimensionless) population densities replace concentrations as the state
  variables in the continuum limit with no change in the mathematical form of
  the governing equations.}\BibitemShut {Stop}%
\bibitem [{\citenamefont {Gillespie}(1977)}]{gillespie1977exact}%
  \BibitemOpen
  \bibfield  {author} {\bibinfo {author} {\bibfnamefont {D.~T.}\ \bibnamefont
  {Gillespie}},\ }\bibfield  {title} {\enquote {\bibinfo {title} {Exact
  stochastic simulation of coupled chemical reactions},}\ }\href@noop {}
  {\bibfield  {journal} {\bibinfo  {journal} {The journal of physical
  chemistry}\ }\textbf {\bibinfo {volume} {81}},\ \bibinfo {pages} {2340--2361}
  (\bibinfo {year} {1977})}\BibitemShut {NoStop}%
\bibitem [{\citenamefont {Cao}, \citenamefont {Gillespie},\ and\ \citenamefont
  {Petzold}(2005)}]{cao2005avoiding}%
  \BibitemOpen
  \bibfield  {author} {\bibinfo {author} {\bibfnamefont {Y.}~\bibnamefont
  {Cao}}, \bibinfo {author} {\bibfnamefont {D.~T.}\ \bibnamefont {Gillespie}},
  \ and\ \bibinfo {author} {\bibfnamefont {L.~R.}\ \bibnamefont {Petzold}},\
  }\bibfield  {title} {\enquote {\bibinfo {title} {Avoiding negative
  populations in explicit {P}oisson tau-leaping},}\ }\href@noop {} {\bibfield
  {journal} {\bibinfo  {journal} {The Journal of chemical physics}\ }\textbf
  {\bibinfo {volume} {123}},\ \bibinfo {pages} {054104} (\bibinfo {year}
  {2005})}\BibitemShut {NoStop}%
\bibitem [{\citenamefont {Cao}, \citenamefont {Gillespie},\ and\ \citenamefont
  {Petzold}(2006)}]{cao2006efficient}%
  \BibitemOpen
  \bibfield  {author} {\bibinfo {author} {\bibfnamefont {Y.}~\bibnamefont
  {Cao}}, \bibinfo {author} {\bibfnamefont {D.~T.}\ \bibnamefont {Gillespie}},
  \ and\ \bibinfo {author} {\bibfnamefont {L.~R.}\ \bibnamefont {Petzold}},\
  }\bibfield  {title} {\enquote {\bibinfo {title} {Efficient step size
  selection for the tau-leaping simulation method},}\ }\href@noop {} {\bibfield
   {journal} {\bibinfo  {journal} {The Journal of chemical physics}\ }\textbf
  {\bibinfo {volume} {124}},\ \bibinfo {pages} {044109} (\bibinfo {year}
  {2006})}\BibitemShut {NoStop}%
\bibitem [{\citenamefont {Kramers}(1940)}]{kramers1940brownian}%
  \BibitemOpen
  \bibfield  {author} {\bibinfo {author} {\bibfnamefont {H.~A.}\ \bibnamefont
  {Kramers}},\ }\bibfield  {title} {\enquote {\bibinfo {title} {Brownian motion
  in a field of force and the diffusion model of chemical reactions},}\
  }\href@noop {} {\bibfield  {journal} {\bibinfo  {journal} {Physica}\ }\textbf
  {\bibinfo {volume} {7}},\ \bibinfo {pages} {284--304} (\bibinfo {year}
  {1940})}\BibitemShut {NoStop}%
\bibitem [{\citenamefont {Moyal}(1949)}]{moyal1949stochastic}%
  \BibitemOpen
  \bibfield  {author} {\bibinfo {author} {\bibfnamefont {J.}~\bibnamefont
  {Moyal}},\ }\bibfield  {title} {\enquote {\bibinfo {title} {Stochastic
  processes and statistical physics},}\ }\href@noop {} {\bibfield  {journal}
  {\bibinfo  {journal} {Journal of the Royal Statistical Society. Series B
  (Methodological)}\ }\textbf {\bibinfo {volume} {11}},\ \bibinfo {pages}
  {150--210} (\bibinfo {year} {1949})}\BibitemShut {NoStop}%
\bibitem [{\citenamefont {Gardiner}\ \emph {et~al.}(1985)\citenamefont
  {Gardiner} \emph {et~al.}}]{gardiner1985handbook}%
  \BibitemOpen
  \bibfield  {author} {\bibinfo {author} {\bibfnamefont {C.~W.}\ \bibnamefont
  {Gardiner}} \emph {et~al.},\ }\href@noop {} {\emph {\bibinfo {title}
  {Handbook of stochastic methods}}},\ Vol.~\bibinfo {volume} {3}\ (\bibinfo
  {publisher} {Springer Berlin},\ \bibinfo {year} {1985})\BibitemShut {NoStop}%
\bibitem [{\citenamefont {Doering}, \citenamefont {Sargsyan},\ and\
  \citenamefont {Sander}(2005)}]{doering2005extinction}%
  \BibitemOpen
  \bibfield  {author} {\bibinfo {author} {\bibfnamefont {C.~R.}\ \bibnamefont
  {Doering}}, \bibinfo {author} {\bibfnamefont {K.~V.}\ \bibnamefont
  {Sargsyan}}, \ and\ \bibinfo {author} {\bibfnamefont {L.~M.}\ \bibnamefont
  {Sander}},\ }\bibfield  {title} {\enquote {\bibinfo {title} {Extinction times
  for birth-death processes: Exact results, continuum asymptotics, and the
  failure of the {Fokker}--{Planck} approximation},}\ }\href@noop {} {\bibfield
   {journal} {\bibinfo  {journal} {Multiscale Modeling \& Simulation}\ }\textbf
  {\bibinfo {volume} {3}},\ \bibinfo {pages} {283--299} (\bibinfo {year}
  {2005})}\BibitemShut {NoStop}%
\bibitem [{\citenamefont {Lin}, \citenamefont {Kim},\ and\ \citenamefont
  {Doering}(2012)}]{lin2012features}%
  \BibitemOpen
  \bibfield  {author} {\bibinfo {author} {\bibfnamefont {Y.~T.}\ \bibnamefont
  {Lin}}, \bibinfo {author} {\bibfnamefont {H.}~\bibnamefont {Kim}}, \ and\
  \bibinfo {author} {\bibfnamefont {C.~R.}\ \bibnamefont {Doering}},\
  }\bibfield  {title} {\enquote {\bibinfo {title} {Features of fast living: on
  the weak selection for longevity in degenerate birth-death processes},}\
  }\href@noop {} {\bibfield  {journal} {\bibinfo  {journal} {Journal of
  Statistical Physics}\ }\textbf {\bibinfo {volume} {148}},\ \bibinfo {pages}
  {647--663} (\bibinfo {year} {2012})}\BibitemShut {NoStop}%
\bibitem [{\citenamefont {Lin}, \citenamefont {Kim},\ and\ \citenamefont
  {Doering}(2015{\natexlab{a}})}]{lin2015demographic}%
  \BibitemOpen
  \bibfield  {author} {\bibinfo {author} {\bibfnamefont {Y.~T.}\ \bibnamefont
  {Lin}}, \bibinfo {author} {\bibfnamefont {H.}~\bibnamefont {Kim}}, \ and\
  \bibinfo {author} {\bibfnamefont {C.~R.}\ \bibnamefont {Doering}},\
  }\bibfield  {title} {\enquote {\bibinfo {title} {Demographic stochasticity
  and evolution of dispersion {I}. {Spatially} homogeneous environments},}\
  }\href@noop {} {\bibfield  {journal} {\bibinfo  {journal} {Journal of
  mathematical biology}\ }\textbf {\bibinfo {volume} {70}},\ \bibinfo {pages}
  {647--678} (\bibinfo {year} {2015}{\natexlab{a}})}\BibitemShut {NoStop}%
\bibitem [{\citenamefont {Lin}, \citenamefont {Kim},\ and\ \citenamefont
  {Doering}(2015{\natexlab{b}})}]{lin2015demographic2}%
  \BibitemOpen
  \bibfield  {author} {\bibinfo {author} {\bibfnamefont {Y.~T.}\ \bibnamefont
  {Lin}}, \bibinfo {author} {\bibfnamefont {H.}~\bibnamefont {Kim}}, \ and\
  \bibinfo {author} {\bibfnamefont {C.~R.}\ \bibnamefont {Doering}},\
  }\bibfield  {title} {\enquote {\bibinfo {title} {Demographic stochasticity
  and evolution of dispersion {II}: {Spatially} inhomogeneous environments},}\
  }\href@noop {} {\bibfield  {journal} {\bibinfo  {journal} {Journal of
  mathematical biology}\ }\textbf {\bibinfo {volume} {70}},\ \bibinfo {pages}
  {679--707} (\bibinfo {year} {2015}{\natexlab{b}})}\BibitemShut {NoStop}%
\bibitem [{\citenamefont {Kampen}(1961)}]{kampen1961power}%
  \BibitemOpen
  \bibfield  {author} {\bibinfo {author} {\bibfnamefont {N.~v.}\ \bibnamefont
  {Kampen}},\ }\bibfield  {title} {\enquote {\bibinfo {title} {A power series
  expansion of the master equation},}\ }\href@noop {} {\bibfield  {journal}
  {\bibinfo  {journal} {Canadian Journal of Physics}\ }\textbf {\bibinfo
  {volume} {39}},\ \bibinfo {pages} {551--567} (\bibinfo {year}
  {1961})}\BibitemShut {NoStop}%
\bibitem [{Note2()}]{Note2}%
  \BibitemOpen
  \bibinfo {note} {The symmetry factor is not strictly required in the context
  of ODE modeling of chemical kinetics (because a constant times a constant is
  still a constant), but it emerges from the CME description of chemical
  kinetics.}\BibitemShut {Stop}%
\bibitem [{Note3()}]{Note3}%
  \BibitemOpen
  \bibinfo {note} {The limit of $h_r\protect \mathaccentV {bar}016{\kappa
  }_r\Delta t$, evaluated at time $t$, as $\Delta t \rightarrow 0$ equals the
  probability that a reaction $r$ takes place somewhere in the system within a
  time window of $t$ to $t+\Delta t$. A good approximation of the probability
  is obtained with finite $\Delta t$ so long as $\Delta t$ is small enough such
  that the probability of two or more reactions of any kind occurring within
  the time window $[t,t+\Delta t)$ is much smaller than the probability of just
  one reaction.}\BibitemShut {Stop}%
\bibitem [{\citenamefont {Schwartz}(2008)}]{schwartz2008biological}%
  \BibitemOpen
  \bibfield  {author} {\bibinfo {author} {\bibfnamefont {R.}~\bibnamefont
  {Schwartz}},\ }\href@noop {} {\emph {\bibinfo {title} {Biological modeling
  and simulation: a survey of practical models, algorithms, and numerical
  methods}}}\ (\bibinfo  {publisher} {MIT Press},\ \bibinfo {year}
  {2008})\BibitemShut {NoStop}%
\bibitem [{\citenamefont {Lin}(2019{\natexlab{a}})}]{ERK_BNGL}%
  \BibitemOpen
  \bibfield  {author} {\bibinfo {author} {\bibfnamefont {Y.~T.}\ \bibnamefont
  {Lin}},\ }\href@noop {} {\enquote {\bibinfo {title} {A {BioNetGen} input file
  for the {ERK} activation model},}\ }\bibinfo {howpublished}
  {\url{https://github.com/RuleWorld/RuleHub/blob/master/Published/Lin2019/ERK_model.bngl}}
  (\bibinfo {year} {\red{accessed May 8, 2019}}{\natexlab{a}})\BibitemShut
  {NoStop}%
\bibitem [{\citenamefont {Lin}(2019{\natexlab{b}})}]{Prion_BNGL}%
  \BibitemOpen
  \bibfield  {author} {\bibinfo {author} {\bibfnamefont {Y.~T.}\ \bibnamefont
  {Lin}},\ }\href@noop {} {\enquote {\bibinfo {title} {A {BioNetGen} input file
  for the prion protein aggregation model.}}\ }\bibinfo {howpublished}
  {\url{https://github.com/RuleWorld/RuleHub/blob/master/Published/Lin2019/prion_model.bngl}}
  (\bibinfo {year} {\red{accessed May 8, 2019}}{\natexlab{b}})\BibitemShut
  {NoStop}%
\bibitem [{\citenamefont {{Lin, Y T}}(2019)}]{TCR_BNGL}%
  \BibitemOpen
  \bibfield  {author} {\bibinfo {author} {\bibnamefont {{Lin, Y T}}},\
  }\href@noop {} {\enquote {\bibinfo {title} {A {BioNetGen} input file for the
  {TCR} signaling model.}}\ }\bibinfo {howpublished}
  {\url{https://github.com/RuleWorld/RuleHub/blob/master/Published/Lin2019/TCR_model.bngl}}
  (\bibinfo {year} {\red{accessed May 8, 2019}})\BibitemShut {NoStop}%
\bibitem [{Note4()}]{Note4}%
  \BibitemOpen
  \bibinfo {note} {Terms higher than second order are dropped because of the
  Pawula theorem\cite {risken1996fokker}, which states that any higher-order
  truncation ($\protect \mathcal {O}\left (1/\Omega ^2\right )$) fails to
  preserve the positivity of $\rho $.}\BibitemShut {Stop}%
\bibitem [{Note5()}]{Note5}%
  \BibitemOpen
  \bibinfo {note} {{\protect \color {red}{Let us use $XY$ to denote the
  population change of a reaction event (in a partially scaled simulation),
  where $X$ is a possibly random multiplier and $Y\sim \protect \text
  {Poisson}\left (\lambda _r \protect \mathaccentV {bar}016{\kappa }_r h_r
  \Delta t \right )$. If the multiplier is distributed such that $\protect
  \mathbb {E}\left [X\right ] = 1/\lambda _r$ and $X$ and $Y$ are independent,
  $\protect \mathbb {E}\left [XY\right ]=h_r\protect \mathaccentV
  {bar}016{\kappa }_r \Delta t$ (as desired for preservation of first moments)
  and $\protect \text {var} \left [ XY \right ]= \protect \text {var} \left [ X
  \right ] \lambda _r h_r \protect \mathaccentV {bar}016{\kappa }_r \Delta t
  \left (1 + \lambda _r h_r \protect \mathaccentV {bar}016{\kappa }_r \Delta t
  \right ) + h_r \protect \mathaccentV {bar}016{\kappa }_r \Delta t /\lambda
  _r$. Because all terms in this expression are positive, it follows that
  $\protect \text {var} \left [ XY \right ]$ is least when $\protect \text
  {var} \left [ X \right ]=0$, i.e., when $X$ is a deterministic variable.
  Recall that for two independently distributed random variables $X$ and $Y$,
  $\protect \mathbb {E}\left [XY\right ] =\protect \mathbb {E}\left [X\right
  ]\protect \mathbb {E}\left [Y\right ]$ and $\protect \text {var} \left [ XY
  \right ] = \protect \text {var} \left [ X \right ]\protect \tmspace
  +\thinmuskip {.1667em} \protect \text {var} \left [ Y \right ] + \protect
  \text {var} \left [ X \right ] \protect \tmspace +\thinmuskip {.1667em}
  \protect \mathbb {E}^2\left [Y\right ] + \protect \mathbb {E}^2\left [X\right
  ]\protect \tmspace +\thinmuskip {.1667em}\protect \text {var} \left [ Y
  \right ]$.}}}\BibitemShut {Stop}%
\bibitem [{\citenamefont {Gibson}\ and\ \citenamefont
  {Bruck}(2000)}]{gibson2000efficient}%
  \BibitemOpen
  \bibfield  {author} {\bibinfo {author} {\bibfnamefont {M.~A.}\ \bibnamefont
  {Gibson}}\ and\ \bibinfo {author} {\bibfnamefont {J.}~\bibnamefont {Bruck}},\
  }\bibfield  {title} {\enquote {\bibinfo {title} {Efficient exact stochastic
  simulation of chemical systems with many species and many channels},}\
  }\href@noop {} {\bibfield  {journal} {\bibinfo  {journal} {The journal of
  physical chemistry A}\ }\textbf {\bibinfo {volume} {104}},\ \bibinfo {pages}
  {1876--1889} (\bibinfo {year} {2000})}\BibitemShut {NoStop}%
\bibitem [{\citenamefont {Walczak}, \citenamefont {Sasai},\ and\ \citenamefont
  {Wolynes}(2005)}]{walczak2005self}%
  \BibitemOpen
  \bibfield  {author} {\bibinfo {author} {\bibfnamefont {A.~M.}\ \bibnamefont
  {Walczak}}, \bibinfo {author} {\bibfnamefont {M.}~\bibnamefont {Sasai}}, \
  and\ \bibinfo {author} {\bibfnamefont {P.~G.}\ \bibnamefont {Wolynes}},\
  }\bibfield  {title} {\enquote {\bibinfo {title} {Self-consistent proteomic
  field theory of stochastic gene switches},}\ }\href@noop {} {\bibfield
  {journal} {\bibinfo  {journal} {Biophysical Journal}\ }\textbf {\bibinfo
  {volume} {88}},\ \bibinfo {pages} {828--850} (\bibinfo {year}
  {2005})}\BibitemShut {NoStop}%
\bibitem [{\citenamefont {Roma}\ \emph {et~al.}(2005)\citenamefont {Roma},
  \citenamefont {O’Flanagan}, \citenamefont {Ruckenstein}, \citenamefont
  {Sengupta},\ and\ \citenamefont {Mukhopadhyay}}]{roma2005optimal}%
  \BibitemOpen
  \bibfield  {author} {\bibinfo {author} {\bibfnamefont {D.~M.}\ \bibnamefont
  {Roma}}, \bibinfo {author} {\bibfnamefont {R.~A.}\ \bibnamefont
  {O’Flanagan}}, \bibinfo {author} {\bibfnamefont {A.~E.}\ \bibnamefont
  {Ruckenstein}}, \bibinfo {author} {\bibfnamefont {A.~M.}\ \bibnamefont
  {Sengupta}}, \ and\ \bibinfo {author} {\bibfnamefont {R.}~\bibnamefont
  {Mukhopadhyay}},\ }\bibfield  {title} {\enquote {\bibinfo {title} {Optimal
  path to epigenetic switching},}\ }\href@noop {} {\bibfield  {journal}
  {\bibinfo  {journal} {Physical Review E}\ }\textbf {\bibinfo {volume} {71}},\
  \bibinfo {pages} {011902} (\bibinfo {year} {2005})}\BibitemShut {NoStop}%
\bibitem [{\citenamefont {Warren}\ and\ \citenamefont {ten
  Wolde}(2005)}]{warren2005chemical}%
  \BibitemOpen
  \bibfield  {author} {\bibinfo {author} {\bibfnamefont {P.~B.}\ \bibnamefont
  {Warren}}\ and\ \bibinfo {author} {\bibfnamefont {P.~R.}\ \bibnamefont {ten
  Wolde}},\ }\bibfield  {title} {\enquote {\bibinfo {title} {Chemical models of
  genetic toggle switches},}\ }\href@noop {} {\bibfield  {journal} {\bibinfo
  {journal} {The Journal of Physical Chemistry B}\ }\textbf {\bibinfo {volume}
  {109}},\ \bibinfo {pages} {6812--6823} (\bibinfo {year} {2005})}\BibitemShut
  {NoStop}%
\bibitem [{\citenamefont {Assaf}, \citenamefont {Roberts},\ and\ \citenamefont
  {Luthey-Schulten}(2011)}]{assaf2011determining}%
  \BibitemOpen
  \bibfield  {author} {\bibinfo {author} {\bibfnamefont {M.}~\bibnamefont
  {Assaf}}, \bibinfo {author} {\bibfnamefont {E.}~\bibnamefont {Roberts}}, \
  and\ \bibinfo {author} {\bibfnamefont {Z.}~\bibnamefont {Luthey-Schulten}},\
  }\bibfield  {title} {\enquote {\bibinfo {title} {Determining the stability of
  genetic switches: explicitly accounting for {mRNA} noise},}\ }\href@noop {}
  {\bibfield  {journal} {\bibinfo  {journal} {Physical review letters}\
  }\textbf {\bibinfo {volume} {106}},\ \bibinfo {pages} {248102} (\bibinfo
  {year} {2011})}\BibitemShut {NoStop}%
\bibitem [{\citenamefont {Strasser}, \citenamefont {Theis},\ and\ \citenamefont
  {Marr}(2012)}]{strasser2012stability}%
  \BibitemOpen
  \bibfield  {author} {\bibinfo {author} {\bibfnamefont {M.}~\bibnamefont
  {Strasser}}, \bibinfo {author} {\bibfnamefont {F.~J.}\ \bibnamefont {Theis}},
  \ and\ \bibinfo {author} {\bibfnamefont {C.}~\bibnamefont {Marr}},\
  }\bibfield  {title} {\enquote {\bibinfo {title} {Stability and multiattractor
  dynamics of a toggle switch based on a two-stage model of stochastic gene
  expression},}\ }\href@noop {} {\bibfield  {journal} {\bibinfo  {journal}
  {Biophysical journal}\ }\textbf {\bibinfo {volume} {102}},\ \bibinfo {pages}
  {19--29} (\bibinfo {year} {2012})}\BibitemShut {NoStop}%
\bibitem [{\citenamefont {Lu}\ \emph {et~al.}(2013)\citenamefont {Lu},
  \citenamefont {Jolly}, \citenamefont {Gomoto}, \citenamefont {Huang},
  \citenamefont {Onuchic},\ and\ \citenamefont
  {Ben-Jacob}}]{lu2013tristability}%
  \BibitemOpen
  \bibfield  {author} {\bibinfo {author} {\bibfnamefont {M.}~\bibnamefont
  {Lu}}, \bibinfo {author} {\bibfnamefont {M.~K.}\ \bibnamefont {Jolly}},
  \bibinfo {author} {\bibfnamefont {R.}~\bibnamefont {Gomoto}}, \bibinfo
  {author} {\bibfnamefont {B.}~\bibnamefont {Huang}}, \bibinfo {author}
  {\bibfnamefont {J.}~\bibnamefont {Onuchic}}, \ and\ \bibinfo {author}
  {\bibfnamefont {E.}~\bibnamefont {Ben-Jacob}},\ }\bibfield  {title} {\enquote
  {\bibinfo {title} {Tristability in cancer-associated {microRNA-TF} chimera
  toggle switch},}\ }\href@noop {} {\bibfield  {journal} {\bibinfo  {journal}
  {The journal of physical chemistry B}\ }\textbf {\bibinfo {volume} {117}},\
  \bibinfo {pages} {13164--13174} (\bibinfo {year} {2013})}\BibitemShut
  {NoStop}%
\bibitem [{\citenamefont {Lok}\ and\ \citenamefont {Brent}(2005)}]{Lok2005}%
  \BibitemOpen
  \bibfield  {author} {\bibinfo {author} {\bibfnamefont {L.}~\bibnamefont
  {Lok}}\ and\ \bibinfo {author} {\bibfnamefont {R.}~\bibnamefont {Brent}},\
  }\bibfield  {title} {\enquote {\bibinfo {title} {Automatic generation of
  cellular reaction networks with {Moleculizer} 1.0},}\ }\href@noop {}
  {\bibfield  {journal} {\bibinfo  {journal} {Nat Biotechnol}\ }\textbf
  {\bibinfo {volume} {23}},\ \bibinfo {pages} {131--136} (\bibinfo {year}
  {2005})}\BibitemShut {NoStop}%
\bibitem [{\citenamefont {Faeder}\ \emph {et~al.}(2005)\citenamefont {Faeder},
  \citenamefont {Blinov}, \citenamefont {Goldstein},\ and\ \citenamefont
  {Hlavacek}}]{Faeder2005}%
  \BibitemOpen
  \bibfield  {author} {\bibinfo {author} {\bibfnamefont {J.~R.}\ \bibnamefont
  {Faeder}}, \bibinfo {author} {\bibfnamefont {M.~L.}\ \bibnamefont {Blinov}},
  \bibinfo {author} {\bibfnamefont {B.}~\bibnamefont {Goldstein}}, \ and\
  \bibinfo {author} {\bibfnamefont {W.~S.}\ \bibnamefont {Hlavacek}},\
  }\bibfield  {title} {\enquote {\bibinfo {title} {Rule-based modeling of
  biochemical networks},}\ }\href@noop {} {\bibfield  {journal} {\bibinfo
  {journal} {Complexity}\ }\textbf {\bibinfo {volume} {10}},\ \bibinfo {pages}
  {22--41} (\bibinfo {year} {2005})}\BibitemShut {NoStop}%
\bibitem [{\citenamefont {Hucka}\ \emph {et~al.}(2018)\citenamefont {Hucka},
  \citenamefont {Bergmann}, \citenamefont {Dräger}, \citenamefont {Hoops},
  \citenamefont {Keating}, \citenamefont {Le~Novère}, \citenamefont {Myers},
  \citenamefont {Olivier}, \citenamefont {Sahle}, \citenamefont {Schaff},
  \citenamefont {Smith}, \citenamefont {Waltemath},\ and\ \citenamefont
  {Wilkinson}}]{Hucka2018}%
  \BibitemOpen
  \bibfield  {author} {\bibinfo {author} {\bibfnamefont {M.}~\bibnamefont
  {Hucka}}, \bibinfo {author} {\bibfnamefont {F.~T.}\ \bibnamefont {Bergmann}},
  \bibinfo {author} {\bibfnamefont {A.}~\bibnamefont {Dräger}}, \bibinfo
  {author} {\bibfnamefont {S.}~\bibnamefont {Hoops}}, \bibinfo {author}
  {\bibfnamefont {S.~M.}\ \bibnamefont {Keating}}, \bibinfo {author}
  {\bibfnamefont {N.}~\bibnamefont {Le~Novère}}, \bibinfo {author}
  {\bibfnamefont {C.~J.}\ \bibnamefont {Myers}}, \bibinfo {author}
  {\bibfnamefont {B.~G.}\ \bibnamefont {Olivier}}, \bibinfo {author}
  {\bibfnamefont {S.}~\bibnamefont {Sahle}}, \bibinfo {author} {\bibfnamefont
  {J.~C.}\ \bibnamefont {Schaff}}, \bibinfo {author} {\bibfnamefont {L.~P.}\
  \bibnamefont {Smith}}, \bibinfo {author} {\bibfnamefont {D.}~\bibnamefont
  {Waltemath}}, \ and\ \bibinfo {author} {\bibfnamefont {D.~J.}\ \bibnamefont
  {Wilkinson}},\ }\bibfield  {title} {\enquote {\bibinfo {title} {The systems
  biology markup language ({SBML}): language specification for level 3 version
  2 core},}\ }\href@noop {} {\bibfield  {journal} {\bibinfo  {journal} {J
  Integr Bioinform}\ }\textbf {\bibinfo {volume} {15}},\ \bibinfo {pages}
  {20170081} (\bibinfo {year} {2018})}\BibitemShut {NoStop}%
\bibitem [{\citenamefont {Chelliah}, \citenamefont {Laibe},\ and\ \citenamefont
  {Le~Novère}(2013)}]{Chelliah2013}%
  \BibitemOpen
  \bibfield  {author} {\bibinfo {author} {\bibfnamefont {V.}~\bibnamefont
  {Chelliah}}, \bibinfo {author} {\bibfnamefont {C.}~\bibnamefont {Laibe}}, \
  and\ \bibinfo {author} {\bibfnamefont {N.}~\bibnamefont {Le~Novère}},\
  }\bibfield  {title} {\enquote {\bibinfo {title} {{BioModels Database}: a
  repository of mathematical models of biological processes},}\ }\href@noop {}
  {\bibfield  {journal} {\bibinfo  {journal} {Methods Mol Biol}\ }\textbf
  {\bibinfo {volume} {1021}},\ \bibinfo {pages} {188--199} (\bibinfo {year}
  {2013})}\BibitemShut {NoStop}%
\bibitem [{\citenamefont {Resat}, \citenamefont {Wiley},\ and\ \citenamefont
  {Dixon}(2001)}]{resat2001probability}%
  \BibitemOpen
  \bibfield  {author} {\bibinfo {author} {\bibfnamefont {H.}~\bibnamefont
  {Resat}}, \bibinfo {author} {\bibfnamefont {H.~S.}\ \bibnamefont {Wiley}}, \
  and\ \bibinfo {author} {\bibfnamefont {D.~A.}\ \bibnamefont {Dixon}},\
  }\bibfield  {title} {\enquote {\bibinfo {title} {Probability-weighted dynamic
  {M}onte {C}arlo method for reaction kinetics simulations},}\ }\href@noop {}
  {\bibfield  {journal} {\bibinfo  {journal} {The Journal of Physical Chemistry
  B}\ }\textbf {\bibinfo {volume} {105}},\ \bibinfo {pages} {11026--11034}
  (\bibinfo {year} {2001})}\BibitemShut {NoStop}%
\bibitem [{\citenamefont {Lin}\ \emph {et~al.}(2018)\citenamefont {Lin},
  \citenamefont {Chylek}, \citenamefont {Lemons},\ and\ \citenamefont
  {Hlavacek}}]{lin2018using}%
  \BibitemOpen
  \bibfield  {author} {\bibinfo {author} {\bibfnamefont {Y.}~\bibnamefont
  {Lin}}, \bibinfo {author} {\bibfnamefont {L.~A.}\ \bibnamefont {Chylek}},
  \bibinfo {author} {\bibfnamefont {N.~W.}\ \bibnamefont {Lemons}}, \ and\
  \bibinfo {author} {\bibfnamefont {W.~S.}\ \bibnamefont {Hlavacek}},\
  }\bibfield  {title} {\enquote {\bibinfo {title} {Using equation-free
  computation to accelerate network-free stochastic simulation of chemical
  kinetics},}\ }\href@noop {} {\bibfield  {journal} {\bibinfo  {journal} {J
  Phys Chem B}\ }\textbf {\bibinfo {volume} {122}},\ \bibinfo {pages}
  {6351--6356} (\bibinfo {year} {2018})}\BibitemShut {NoStop}%
\bibitem [{\citenamefont {Thomas}\ \emph {et~al.}(2016)\citenamefont {Thomas},
  \citenamefont {Chylek}, \citenamefont {Colvin}, \citenamefont {Sirimulla},
  \citenamefont {Clayton}, \citenamefont {Hlavacek},\ and\ \citenamefont
  {Posner}}]{Thomas2016}%
  \BibitemOpen
  \bibfield  {author} {\bibinfo {author} {\bibfnamefont {B.~R.}\ \bibnamefont
  {Thomas}}, \bibinfo {author} {\bibfnamefont {L.~A.}\ \bibnamefont {Chylek}},
  \bibinfo {author} {\bibfnamefont {J.}~\bibnamefont {Colvin}}, \bibinfo
  {author} {\bibfnamefont {S.}~\bibnamefont {Sirimulla}}, \bibinfo {author}
  {\bibfnamefont {A.~H.~A.}\ \bibnamefont {Clayton}}, \bibinfo {author}
  {\bibfnamefont {W.~S.}\ \bibnamefont {Hlavacek}}, \ and\ \bibinfo {author}
  {\bibfnamefont {R.~G.}\ \bibnamefont {Posner}},\ }\bibfield  {title}
  {\enquote {\bibinfo {title} {{BioNetFit}: a fitting tool compatible with
  {BioNetGen}, {NFsim} and distributed computing environments},}\ }\href@noop
  {} {\bibfield  {journal} {\bibinfo  {journal} {Bioinformatics}\ }\textbf
  {\bibinfo {volume} {32}},\ \bibinfo {pages} {798--800} (\bibinfo {year}
  {2016})}\BibitemShut {NoStop}%
\bibitem [{\citenamefont {Chylek}\ \emph
  {et~al.}(2014{\natexlab{b}})\citenamefont {Chylek}, \citenamefont {Akimov},
  \citenamefont {Dengjel}, \citenamefont {Rigbolt}, \citenamefont {Hu},
  \citenamefont {Hlavacek},\ and\ \citenamefont {Blagoev}}]{Chylek2014TCR}%
  \BibitemOpen
  \bibfield  {author} {\bibinfo {author} {\bibfnamefont {L.~A.}\ \bibnamefont
  {Chylek}}, \bibinfo {author} {\bibfnamefont {V.}~\bibnamefont {Akimov}},
  \bibinfo {author} {\bibfnamefont {J.}~\bibnamefont {Dengjel}}, \bibinfo
  {author} {\bibfnamefont {K.~T.~G.}\ \bibnamefont {Rigbolt}}, \bibinfo
  {author} {\bibfnamefont {B.}~\bibnamefont {Hu}}, \bibinfo {author}
  {\bibfnamefont {W.~S.}\ \bibnamefont {Hlavacek}}, \ and\ \bibinfo {author}
  {\bibfnamefont {B.}~\bibnamefont {Blagoev}},\ }\bibfield  {title} {\enquote
  {\bibinfo {title} {Phosphorylation site dynamics of early {T}-cell receptor
  signaling},}\ }\href@noop {} {\bibfield  {journal} {\bibinfo  {journal} {PLOS
  ONE}\ }\textbf {\bibinfo {volume} {9}},\ \bibinfo {pages} {e104240} (\bibinfo
  {year} {2014}{\natexlab{b}})}\BibitemShut {NoStop}%
\bibitem [{\citenamefont {Liu}\ \emph {et~al.}(2010)\citenamefont {Liu},
  \citenamefont {Mobassera}, \citenamefont {Shaffer}, \citenamefont {Watson},\
  and\ \citenamefont {Cao}}]{Liu2010multistate}%
  \BibitemOpen
  \bibfield  {author} {\bibinfo {author} {\bibfnamefont {Z.}~\bibnamefont
  {Liu}}, \bibinfo {author} {\bibfnamefont {U.~J.}\ \bibnamefont {Mobassera}},
  \bibinfo {author} {\bibfnamefont {C.~A.}\ \bibnamefont {Shaffer}}, \bibinfo
  {author} {\bibfnamefont {L.~T.}\ \bibnamefont {Watson}}, \ and\ \bibinfo
  {author} {\bibfnamefont {Y.}~\bibnamefont {Cao}},\ }\bibfield  {title}
  {\enquote {\bibinfo {title} {Multistate modeling and simulation for
  regulatory networks},}\ }in\ \href@noop {} {\emph {\bibinfo {booktitle}
  {Proceedings of the 2010 Winter Simulation Conference}}},\ \bibinfo {editor}
  {edited by\ \bibinfo {editor} {\bibfnamefont {B.}~\bibnamefont {Johansson}},
  \bibinfo {editor} {\bibfnamefont {S.}~\bibnamefont {Jain}}, \bibinfo {editor}
  {\bibfnamefont {J.}~\bibnamefont {Montoya-Torres}}, \bibinfo {editor}
  {\bibfnamefont {J.}~\bibnamefont {Hugan}}, \ and\ \bibinfo {editor}
  {\bibfnamefont {E.}~\bibnamefont {Y\"{u}cesan}}}\ (\bibinfo {year} {2010})\
  pp.\ \bibinfo {pages} {631--642}\BibitemShut {NoStop}%
\bibitem [{\citenamefont {Van~Valen}(1974)}]{van1974multivariate}%
  \BibitemOpen
  \bibfield  {author} {\bibinfo {author} {\bibfnamefont {L.}~\bibnamefont
  {Van~Valen}},\ }\bibfield  {title} {\enquote {\bibinfo {title} {Multivariate
  structural statistics in natural history},}\ }\href@noop {} {\bibfield
  {journal} {\bibinfo  {journal} {Journal of Theoretical Biology}\ }\textbf
  {\bibinfo {volume} {45}},\ \bibinfo {pages} {235--247} (\bibinfo {year}
  {1974})}\BibitemShut {NoStop}%
\end{thebibliography}
\end{document}